\documentclass{article}

\newif\ifcsur\csurfalse

\ifcsur
\else
\fi

\ifcsur\else
\def\citeyear#1{\cite{#1}}
\newenvironment{describe}[1]{\begin{description}}{\end{description}}
\def\tbl#1{\caption{#1}}
\fi

\ifcsur
\acmVolume{46}
\acmNumber{3}
\acmArticle{34}
\acmYear{2014}
\acmMonth{1}
\doi{2506375}
\fi

\usepackage{hyperref}
\usepackage{triotex}

\usepackage{tikz}
\usetikzlibrary{shapes,arrows,positioning,fit,backgrounds,calc}

\usepackage{subfigure}

\usepackage{chngpage}

\usepackage{amsfonts}
\usepackage{amsmath}
\usepackage{amssymb}

\usepackage{array}
\usepackage{url}
\usepackage{xspace}

\usepackage{graphicx}

\usepackage{eiffel}

\newif\ifdraft\draftfalse

\ifdraft
 \newcommand\remk[1]{\mymarginpar{\raggedright\hbadness=10000\tiny\it #1\par}}
\else
   \newcommand\remk[1]    {}
\fi

\ifdraft\overfullrule3pt\fi

\newcommand{\limpl}{\mathop{\Longrightarrow}}
\newcommand{\liff}{\mathop{\Longleftrightarrow}}

\newcommand{\ecfootnote}[1]{\mbox{\lstinline[basicstyle=\footnotesize]|#1|}}
\newcommand{\ecplain}[1]{\lstinline[basicstyle=\normalsize]|#1|}
\newcommand{\ec}[1]{\mbox{\ecplain{#1}}}
\lstnewenvironment{eifeq}[1][3cm]
{\lstset{language=OOSC2Eiffel,xleftmargin=#1,numbers=none,basicstyle=\normalsize}}
{}

\newcommand{\ginpink}{\textsf{gin-pink}\xspace}

\date{}

\ifcsur\fi
\title{Loop invariants: analysis, classification, and examples}

\ifcsur
\author{CARLO A. FURIA \affil{ETH Zurich} 
        BERTRAND MEYER \affil{ETH Zurich, ITMO St. Petersburg, and Eiffel Software}
        SERGEY VELDER \affil{ITMO St. Petersburg}}
\else
\author{Carlo A. Furia $\cdot$ Bertrand Meyer $\cdot$ Sergey Velder}
\fi

\ifcsur
\begin{abstract}
  Software verification has emerged as a key concern for ensuring the
  continued progress of information technology. Full verification
  generally requires, as a crucial step, equipping each loop with a
  ``loop invariant''. Beyond their role in verification, loop
  invariants help program understanding by providing fundamental
  insights into the nature of algorithms.  In practice, finding sound
  and useful invariants remains a challenge. Fortunately, many
  invariants seem intuitively to exhibit a common
  flavor. Understanding these fundamental invariant patterns could
  therefore provide help for understanding and verifying a large
  variety of programs. 

  We performed a systematic identification, validation, and
  classification of loop invariants over a range of fundamental
  algorithms from diverse areas of computer science. This article
  analyzes the patterns, as uncovered in this study, governing how
  invariants are derived from postconditions; it proposes a taxonomy
  of invariants according to these patterns, and presents its
  application to the algorithms reviewed. The discussion also shows
  the need for high-level specifications based on ``domain
  theory''. It describes how the invariants and the corresponding
  algorithms have been mechanically verified using an automated
  program prover; the proof source files are available.  The
  contributions also include suggestions for invariant inference and
  for model-based specification.
\end{abstract}

\fi

\ifcsur
\acmformat{Carlo A. Furia, Bertrand Meyer, and Sergey Velder. 2014. Loop invariants: Analysis, classification, and examples.}
\fi

\begin{document}

\maketitle

\ifcsur\else
\begin{abstract}
  Software verification has emerged as a key concern for ensuring the
  continued progress of information technology. Full verification
  generally requires, as a crucial step, equipping each loop with a
  ``loop invariant''. Beyond their role in verification, loop
  invariants help program understanding by providing fundamental
  insights into the nature of algorithms.  In practice, finding sound
  and useful invariants remains a challenge. Fortunately, many
  invariants seem intuitively to exhibit a common
  flavor. Understanding these fundamental invariant patterns could
  therefore provide help for understanding and verifying a large
  variety of programs. 

  We performed a systematic identification, validation, and
  classification of loop invariants over a range of fundamental
  algorithms from diverse areas of computer science. This article
  analyzes the patterns, as uncovered in this study, governing how
  invariants are derived from postconditions; it proposes a taxonomy
  of invariants according to these patterns, and presents its
  application to the algorithms reviewed. The discussion also shows
  the need for high-level specifications based on ``domain
  theory''. It describes how the invariants and the corresponding
  algorithms have been mechanically verified using an automated
  program prover; the proof source files are available.  The
  contributions also include suggestions for invariant inference and
  for model-based specification.
\end{abstract}


\fi

\ifcsur
\else
\newpage
\tableofcontents
\newpage
\fi


\section{Introduction: inductive invariants} \label{sec:intro}

The problem of guaranteeing program correctness remains one of the
central challenges of software engineering, of considerable importance
to the information technology industry and to society at large, which
increasingly depends, for almost all of its processes, on correctly
functioning programs. As defined by Tony Hoare~\citeyear{Hoare03}, the
``Grand Challenge of Program Verification'' mobilizes many researchers
and practitioners using a variety of techniques.

Some of these techniques, such as model checking~\cite{MCbook} and
abstract interpretation~\cite{CC77}, are directed at finding specific
errors, such as the possible violation of a safety property. An
advantage of these techniques is that they work on programs as they
are, without imposing a significant extra annotation effort on
programmers. For full functional correctness---the task of proving
that a program satisfies a complete specification---the approach of
choice remains, for imperative programs, the Floyd-Hoare-Dijkstra
style of axiomatic semantics. In this approach, programs must be
equipped with annotations in the form of \emph{assertions}. Every
loop, in particular, must have a \emph{loop invariant}.


Finding suitable loop invariants is a crucial and delicate step to
verification. Although some programmers may see invariant elicitation
as a chore needed only for formal verification, the concept is in fact
widely useful, including for informal development: the invariant gives
fundamental information about a loop, showing what it is trying to
achieve and how it achieves it, to the point that (in some people's
view at least) it is impossible to understand a loop without knowing
its invariant.

To explore and illustrate this view, we have investigated a body
of representative loop algorithms in several areas of computer science,
to identify the corresponding invariants, and found that they follow a
set of standard patterns. We set out to uncover, catalog, classify,
and verify these patterns, and report our findings in the present
article.

Finding an invariant for a loop is traditionally the responsibility of
a human: either the person performing the verification, or the
programmer writing the loop in the first place (a better solution, when
applicable, is the \emph{constructive} approach to programming
advocated by Dijkstra and others~\cite{Dij76,Gri81,BM80}). More
recently, techniques have been developed for \emph{inferring}
invariants automatically, or semi-automatically with some human
help~(we review them in Section~\ref{sec:relat-work:-autom}). We hope
that the results reported here will be useful in both cases: for
humans, to help obtain the loop invariants of new or existing
programs, a task that many programmers still find challenging; and for
invariant inference tools.

For all algorithms presented in the paper\footnote{With the exception
  of those in Sections~\ref{sec:rot-calipers} and~\ref{sec:pagerank},
  whose presentation is at a higher level of abstraction, so that a
  complete formalization would have required complex axiomatization of
  geometric and numerical properties beyond the focus of this paper.},
we wrote fully annotated implementations and processed the result with
the Boogie program verifier~\cite{Lei08-Boogie2}, providing proofs of
correctness. The Boogie implementations are available at:\footnote{In
  the repository, the branch \texttt{inv\_survey} contains only the
  algorithms described in the paper; see \url{http://goo.gl/DsdrV} for
  instruction on how to access it.}
\begin{center}
  \url{http://bitbucket.org/sechairethz/verified/}
\end{center}
This verification result reinforces the confidence in the correctness
of the algorithms presented in the paper and their practical
applicability.

The rest of this introductory section recalls the basic properties of
invariants. Section~\ref{sec:domain-theory} introduces a style of
expressing invariants based on ``domain theory'', which can often be
useful for clarity and
expressiveness. Section~\ref{sec:classification} presents two
independent classifications of loop invariant clauses, according to
their role and syntactic similarity with respect to the postcondition.
Section~\ref{sec:catalog} presents 21 algorithms from various domains;
for each algorithm, it presents an implementation in pseudo-code
annotated with complete specification and loop invariants.
Section~\ref{sec:relat-work:-autom} discusses some related techniques
to infer invariants or other specification elements automatically.
Section~\ref{sec:boogie-assess} draws lessons from the verification effort.
Section~\ref{sec:conclusions}  concludes.

\subsection{Loop invariants basics} \label{sec:invariants-basics}

The loop invariants of the axiomatic approach go back to
Floyd~\citeyear{Flo67} and Hoare~\citeyear{Hoare69} (see Hatcliff et
al.~\citeyear{HatcliffLLMP12} for a survey of notations for and
variants of the fundamental idea). For this approach and for the
present article, a loop invariant is not just a quantity that remains
unchanged throughout executions of the loop body (a notion that has
also been studied in the literature), but more specifically an
``inductive invariant'', of which the precise definition appears next.
Program verification also uses other kinds of invariant, notably class
invariants~\cite{Hoare72,OOSC-2}, which the present discussion surveys
only briefly in Section~\ref{sec:other-kinds}.

The notion of loop invariant is easy to express in the following loop
syntax taken from Eiffel:

\begin{minipage}{0.9\linewidth}
\begin{lstlisting}
    from
       Init
    invariant
       Inv
    until
       Exit
    variant
       Var
    loop
       Body
    end
\end{lstlisting}
\end{minipage} \\
(the \ec{variant} clause helps establish termination as discussed
later). \ec{Init} and \ec{Body} are each a compound (a list of
instructions to be executed in sequence); either or both can be empty,
although \ec{Body} normally will not. \ec{Exit} and \ec{Inv} (the
inductive invariant) are both Boolean expressions, that is to say,
predicates on the program state.  The semantics of the loop is:
\begin{enumerate}
\item \label{loop:init} Execute \ec{Init}.
\item \label{loop:repeat} Then, if \ec{Exit} has value \ec{True}, do
  nothing; if it has value \ec{False}, execute \ec{Body}, and repeat
  step~\ref{loop:repeat}.
\end{enumerate}
Another way of stating this informal specification is that the
execution of the loop body consists of the execution of \ec{Init} followed
by zero or more executions of \ec{Body}, stopping as soon as \ec{Exit} becomes
\ec{True}.

There are many variations of the loop construct in imperative
programming languages: ``while'' forms which use a continuation
condition rather than the inverse exit condition; ``do-until'' forms
that always execute the loop body at least once, testing for the
condition at the end rather than on entry; ``for'' or ``do'' forms
(``\ec{across}'' in Eiffel) which iterate over an integer interval or
a data structure. They can all be derived in a straightforward way
from the above basic form, on which we will rely throughout this
article.

The invariant \ec{Inv} plays no direct role in the informal semantics,
but serves to reason about the loop and its correctness. \ec{Inv} is a
correct invariant for the loop if it satisfies the following
conditions:
\begin{enumerate}
\item Every execution of \ec{Init}, started in the state preceding the
  loop execution, will yield a state in which \ec{Inv} holds.
\item Every execution of \ec{Body}, started in any state in which
  \ec{Inv} holds and \ec{Exit} does not hold, will yield a state in
  which \ec{Inv} holds again.
\end{enumerate}

If these properties hold, then any terminating execution of the loop
will yield a state in which both \ec{Inv} and \ec{Exit} hold. This
result is a consequence of the loop semantics, which defines the loop
execution as the execution of \ec{Init} followed by zero or more
executions of \ec{Body}, each performed in a state where \ec{Exit}
does not hold. If \ec{Init} ensures satisfaction of the invariant, and
any one execution of \ec{Body} preserves it (it is enough to obtain
this property for executions started in a state not satisfying
\ec{Exit}), then \ec{Init} followed by \emph{any} number of executions
of \ec{Body} will.

Formally, the following classic inference rule~\cite{Hoare72,OOSC-2} uses the
invariant to express the correctness requirement on any loop:
\[
\frac
{\text{\ec{\{P\} Init \{Inv\}}},\quad 
 \text{\ec{\{Inv land lnot Exit\} Body \{Inv\}}}}
{\text{\ec{\{P\} from Init until Exit loop Body end \{Inv land Exit\}}}} .
\]
This is a partial correctness rule, useful only for loops that
terminate. Proofs of termination are in general handled separately
through the introduction of a loop variant: a value from a
well-founded set, usually taken to be the set of natural numbers,
which decreases upon each iteration (again, it is enough to show that
it does so for initial states not satisfying \ec{Exit}). Since in a
well-founded set all decreasing sequences are finite, the existence of
a variant expression implies termination. The rest of this discussion
concentrates on the invariants; it only considers terminating
algorithms, of course, and includes the corresponding \ec{variant}
clauses, but does not explain why the corresponding expression are
indeed loop variants (non-negative and decreasing). Invariants,
however, also feature in termination proofs, where they ensure that
the variant ranges over a well-founded set (or, equivalently, the
values it takes are bounded from below).

If a loop is equipped with an invariant, proving its partial
correctness means establishing the two hypotheses in the above rules:
\begin{itemize}
\item \ec{\{P\} Init \{Inv\}}, stating that the initialization ensures the
      invariant, is called the \emph{initiation} property.
    \item \ec{\{Inv land lnot Exit\} Body \{Inv\}}, stating that the
      body preserves the invariant, is called the \emph{consecution}
      (or \emph{inductiveness}) property.
\end{itemize}

\subsection{A constructive view} \label{sec:constr-view}

We may look at the notion of loop invariant from the constructive
perspective of a programmer directing his or her program to reach a
state satisfying a certain desired property, the postcondition. In
this view, program construction is a form of problem-solving, and the
various control structures are problem-solving
techniques~\cite{Dij76,BM80,Gri81,Morgan94}; a loop solves a problem
through successive approximation.

\begin{figure}[!hbt]
  \centering
\begin{tikzpicture}
\def\postC{(3.8,0) circle [x radius=20mm, y radius=15mm]}
\def\exitC{(2.6,0) circle [x radius=5mm, y radius=35mm]}
\def\invC{(-1,0) circle [x radius=45mm, y radius=15mm]}

\begin{scope}[fill opacity=0.5]
\fill[red!30] \postC; 
\fill[yellow!40] \exitC; 
\fill[cyan!30] \invC; 
\end{scope}

\draw[red!50!black!50,ultra thick] \postC node [above right=16mm and 2mm] {\ec{Postcondition}};
\draw[yellow!50!black!50,ultra thick] \exitC node [above=36mm] {\ec{Exit condition}};
\draw[cyan!50!black!50,ultra thick] \invC node [above=16mm] {\ec{Invariant}};

\node [draw=black,circle,fill=black,label=above:\ec{Previous state}] (before) at ($(-1,0)+(-40mm,25mm)$) {};

\node [draw=black!40,circle,fill=black!40] (step1) at ($(-1,0)+(-40mm,0)$) {};
\node [right=28mm of step1,draw=black!40,circle,fill=black!40] (step2) {};
\node [above right=3mm and 18mm of step2,draw=black!40,circle,fill=black!40] (step3) {};
\node [below right=8mm and 9mm of step3,draw=black!40,circle,fill=black!40] (step4) {};
\node [draw=black!40,circle,fill=black!40] (step5) at ($(3.1,0)+(-6mm,0)$) {};

\begin{scope}[->,very thick]
\path (before) edge node [right, near start] {Initialization} (step1);
\path (step1) edge node [above] {Body} (step2);
\path (step2) edge node [above=0mm] {Body} (step3);
\path (step3) edge [dotted] (step4);
\path (step4) edge node [above=1mm, near start] {Body} (step5);
\end{scope}

\end{tikzpicture}
 \caption{The loop as a computation by approximation.}
  \label{fig:loop-as-approximation}
\end{figure}
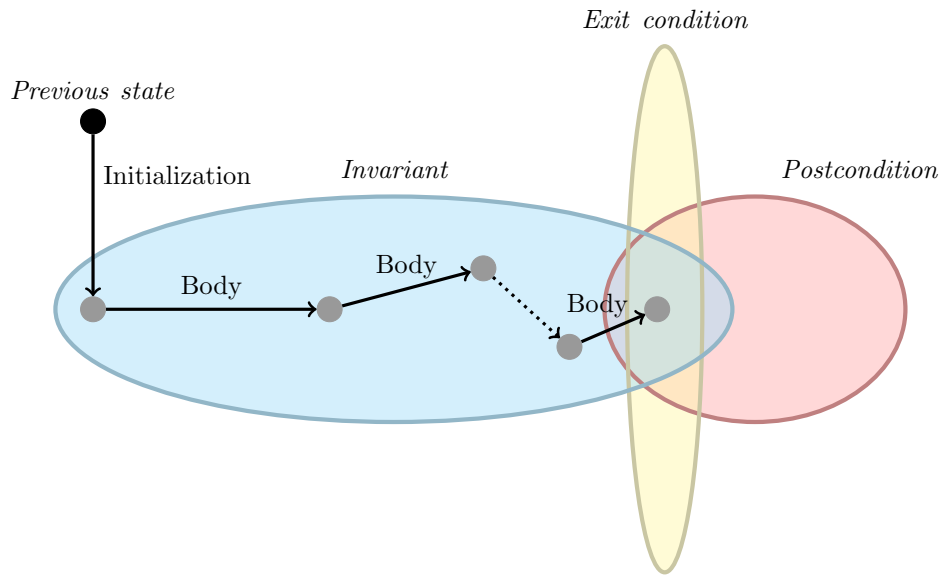

The idea of this solution, illustrated by
Figure~\ref{fig:loop-as-approximation}, is the following:
\begin{itemize}
\item Generalize the postcondition (the characterization of possible
  solutions) into a broader condition: the invariant.
\item As a result, the postcondition can be defined as the combination
  (``and'' in logic, intersection in the figure) of the invariant and
  another condition: the exit condition.
\item Find a way to reach the invariant from the previous state of the
  computation: the initialization.
\item Find a way, given a state that satisfies the invariant, to get
  to another state, still satisfying the invariant but closer, in some
  appropriate sense, to the exit condition: the body.
\end{itemize}
For the solution to reach its goal after a finite number of steps we
need a notion of discrete ``distance'' to the exit condition. This is
the loop variant.

The importance of the above presentation of the loop process is that
it highlights the nature of the invariant: it is a generalized form of
the desired postcondition, which in a special case (represented by the
exit condition) will give us that postcondition. This view of the
invariant, as a particular way of generalizing the desired goal of the
loop computation, explains why the loop invariant is such an important
property of loops; one can argue that understanding a loop means
understanding its invariant (in spite of the obvious observation that
many programmers write loops without ever formally learning the notion
of invariant, although we may claim that if they understand what they
are doing they are relying on some intuitive understanding of the
invariant anyway, like Moli\`ere's Mr.~Jourdain speaking in prose
without knowing it).

The key results of this article can be described as generalization
strategies to obtain invariants from postconditions.

\subsection{A basic example} \label{sec:basic-example}

To illustrate the above ideas, the 2300-year-old example of Euclid's
algorithm, while very simple, is still a model of elegance. The
postcondition of the algorithm is
\[
\ec{Result} = \gcd(a,b) ,
\]
where the positive integers $a$ and $b$ are the input and $\gcd$ is
the mathematical Greatest Common Divisor function. The generalization
is to replace this condition by
\begin{equation}
\ec{Result} = x  \quad\wedge\quad 
\gcd(\ec{Result}, x) = \gcd(a, b)
\label{eq:gcd-relax}
\end{equation}
with a new variable $x$, taking advantage of the mathematical property
that, for every $x$,
\begin{equation}
\gcd (x, x) = x .
\label{eq:gcd-xx}
\end{equation}
The second conjunct, a generalization of the
postcondition, will serve as the invariant; the first conjunct will
serve as the exit condition. To obtain the loop body we take advantage
of another mathematical property: for every $x > y$,
\begin{equation}
\gcd (x, y) = \gcd (x - y, y) ,
\label{eq:gcd-minus}
\end{equation}
yielding the well-known algorithm in
Figure~\ref{fig:gcd-subtract}. (As with any assertion, writing clauses
successively in the invariant is equivalent to a logical conjunction.)
This form of Euclid's algorithm uses subtraction; another form, given
in Section~\ref{sec:GCD-division}, uses integer division.

\begin{figure}[!htb]
\begin{lstlisting}
from
  Result := a ; x := b
invariant
  Result > 0
  x > 0
  $\gcd$ (Result, x) = $\gcd$ (a, b)
until
  Result = x
loop
  if Result > x then
     Result := Result - x
  else     -- Here $x$ is strictly greater than $\textbf{Result}$
     x := x - Result
  end
variant
  $\max$ (Result, x)   
end
\end{lstlisting}
  \caption{Greatest common divisor with substraction.}
  \label{fig:gcd-subtract}
\end{figure}

We may use this example to illustrate some of the orthogonal
categories in the classification developed in the rest of this article:
\begin{itemize}
\item The last clause of the invariant is an \emph{essential
    invariant}, representing a weakening of the postcondition. The
  first two clauses are a \emph{bounding invariant}, indicating that
  the state remains within certain general boundaries, and ensuring
  that the ``essential'' part is defined.
\item The essential invariant is a \emph{conservation invariant},
  indicating that a certain quantity remains equal to its original
  value.
\item The strategy that leads to this conservation invariant is
  \emph{uncoupling}, which replaces a property of one variable
  (\ec{Result}), used in the postcondition, by a property of two
  variables (\ec{Result} and $x$), used in the invariant.
\end{itemize}

The proof of correctness follows directly from the mathematical
property stated: \eqref{eq:gcd-xx} establishes initiation, and
\eqref{eq:gcd-minus} establishes consecution.

Section~\ref{sec:GCD-division} shows how the same technique is
applicable backward, to guess likely loop invariants given an 
algorithm annotated with pre- and postcondition: mutating the latter yields
a suitable loop invariant.

\subsection{Other kinds of invariant} \label{sec:other-kinds}

Loop invariants are the focus of this article, but before we return to
them it is useful to list some other kinds of invariant encountered in
software. (Yet other invariants, which lie even further beyond the
scope of this discussion, play fundamental roles in fields such as
physics; consider for example the invariance of the speed of light
under a Lorentz transformation, and of time under a Galilean
transformation.)

In object-oriented programming, a class invariant (also directly
supported by the Eiffel notation~\cite{eiffel-ECMA}) expresses a
property of a class that:
\begin{itemize}
\item Every instance of the class possesses immediately after
  creation, and
\item Every exported feature (operation) of the class preserves,
\end{itemize}
with the consequence that whenever such an object is accessible to the
rest of the software it satisfies the invariant, since the life of an
object consists of creation followed by any number of ``qualified''
calls \ec{x.f} to exported features \ec{f} by clients of the class.
The two properties listed are strikingly similar to initiation and
consecution for loop invariants, and the connection appears clearly if
we model the life of an object as a loop:

\vspace{2mm}
\begin{lstlisting}
from
  create x.make  -- Written in some languages as $x := \textbf{new}\;C()$
invariant
  CI        -- The class invariant
until
  (*``$x$ is no longer needed''*)
loop
  x.some_feature_of_the_class
end
\end{lstlisting}

Also useful are Lamport-style invariants~\cite{Lam77} used to reason
about concurrent programs, which obviate the need for the ``ghost
variables'' of the Owicki-Gries method~\cite{OG76}). Like other
invariants, a Lamport invariant is a predicate on the program state;
the difference is that the definition of the states involves not only
the values of the program's variables but also the current point of
the execution of the program (``Program Counter'' or PC) and, in the
case of a concurrent program, the collection of the PCs of all its
concurrent processes. An example of application is the answer to the
following problem posed by Lamport~\citeyear{Lamport09}. 
\begin{quote}
Consider $N$ processes numbered from 0 through $N - 1$ in which each process $i$ executes
\begin{align*}
\ell_0^i: & \quad x[i] := 1 \\
\ell_1^i: & \quad y[i] := x[(i - 1) \bmod N] \\
\ell_2^i:
\end{align*}
and stops, where each $x[i]$ initially equals 0. (The reads and writes
of each $x[i]$ are assumed to be atomic.) [\ldots] The algorithm
[\ldots] maintains an inductive invariant. Do you know what that
invariant is?
\end{quote}
If we associate a proposition $@(m,i)$ for $m = 1, 2, 3$ that holds precisely when the execution of process $i$ reaches location $\ell_m^i$, an invariant for the algorithm can be expressed as:
\begin{align*}
@(2, i) & \limpl \left(\begin{array}{c}
     @(0, (i - 1) \bmod N) \land y[i] = 0  \\
     \vee \\
     @(1, (i - 1) \bmod N) \land y[i] = 1 \\
     \vee \\
     @(2, (i - 1) \bmod N) \land y[i] = 1
   \end{array}\right) .
 \end{align*}

Yet another kind of invariant occurs in the study of dynamical
systems, where an invariant is a region $I \subseteq \numset{S}$ of
the state space $\numset{S}$ such that any trajectory starting in $I$
or entering it stays in $I$ indefinitely in the future:
\[
\forall x \in I, \forall t \in \timedomain : \Phi(t, x) \in I ,
\]
where $\timedomain$ is the time domain and $\Phi: \timedomain \times
\numset{S} \to \numset{S} $ is the evolution function. The connection
between dynamical system invariants and loop invariants is clear in
the constructive view (Section~\ref{sec:constr-view}), and can be
formally derived by modeling programs as dynamical systems or using
some other operational formalism~\cite{FMMR-TimeBook-12}.  The
differential invariants introduced in the study of hybrid
systems~\cite{Platzer10} are also variations of the invariants defined
by dynamical systems.

\section{Expressing invariants: domain theory} \label{sec:domain-theory}

To discuss and compare invariants we need to settle on the
expressiveness of the underlying invariant language: what do we accept
as a loop invariant?

The question involves general assertions, not just invariants; more
generally, we must make sure that any convention for invariants is
compatible with the general scheme used for pre/post specification,
since an invariant is a mutation (possibly a weakening) of the
postcondition.

The mathematical answer to the basic question is simple: an assertion
other than a routine postcondition, in particular a loop invariant, is
a \emph{predicate} on the program state. For example, the assertion $x
> 0$, where $x$ is a program variable, is the predicate that holds of
all computation states in which the value of that variable is
positive. (Another view would consider it as the \emph{subset} of the
state space containing all states that satisfy the condition; the two
views are equivalent since the predicate is the characteristic
function of the subset, and the subset is the inverse domain of ``true''
for the predicate.)

A routine postcondition is usually a predicate on \emph{two} states,
since the specification of a routine generally relates new values to
original ones. For example, an increment routine yields a state in
which the counter's value is one more on exit than on entry. The
\ec{old} notation, available for postconditions in Eiffel and other
programming languages supporting contracts, reflects this need; for
example, a postcondition clause could read \ec{counter = old counter +
  1}.  Other notations, notably the Z specification
language~\cite{Z-ref}, have a notation for ``new'' rather than ``old'',
as in \ec{counter' = counter + 1} where the primed variable denotes
the new value. Although invariants are directly related to
postconditions, we will be able in this discussion to avoid such
notations and treat invariants as one-state functions. (Technically, 
this is always possible by recording the entry value as part of the
state.)

Programming languages offer a mechanism directly representing
predicates on states: Boolean expressions. This construct can
therefore be used---as in the $x > 0$ example---to represent
assertions; this is what assertion-aware programming languages
typically do, often extending it with special notations such as
\ec{old} and support for quantifiers.

This basic language decision leaves open the question of the
\emph{level of expressiveness} of assertions. There are two
possibilities:
\begin{itemize}
\item Allow assertions, in particular postconditions and loop
  invariants, to use \emph{functions} and \emph{predicates} defined
  using some appropriate mechanism (often, the programming language's
  function declaration construct) to express high-level properties
  based on a domain theory covering specifics of the
  application area. We call this approach \emph{domain
    theory}.\footnote{No relation with the study of partially ordered
    sets, also called domain theory~\cite{DT-handbook}.}

\item Disallow this possibility, requiring assertions always to be
  expressed in terms of the constructs of the assertion language,
  without functions. We call this approach \emph{atomic assertions}.
\end{itemize}

The example of Euclid's algorithm above, simple as it is, was already
an example of the domain-theory-based approach because of its use of a
function $\gcd$ in the invariant clause
\begin{equation}
 \gcd(\ec{Result}, x) = \gcd(a,b)
 \label{eq:gcd-invariant}
\end{equation}
corresponding to a weakening of the routine postcondition
\begin{equation*}
	\ec{Result} = \gcd(a,b) .
\end{equation*}
 
It is possible to do without such a function by going back to the
basic definition of the greatest common denominator. In such an
atomic-assertion style, the postcondition would read
\begin{center}
\begin{adjustwidth}{-1cm}{-1cm}
\begin{center}
\begin{tabular}[h]{lp{4.3cm}}
  \ec{Result > 0}    & (Alternatively, \ec{Result >= 1}) \\
  \ec{a modOp Result = 0}  & (\ec{Result} divides \ec{a}) \\
  \ec{b modOp Result = 0}  & (\ec{Result} divides \ec{b}) \\
  \ec{$\forall i \in \naturals$: (a modOp i = 0) land (b modOp i = 0) $\;$implies$\;$ i <= Result}
 & (\ec{Result} is the greatest of all the numbers that satisfy the preceding properties).
\end{tabular}
\end{center}
\end{adjustwidth}
\end{center}
Expressing the invariant in the same style requires several more lines
since the definition of the greatest common divisor must be expanded
for both sides of~\eqref{eq:gcd-invariant}.

Even for such a simple example, the limitations of the
atomic-assertion style are clear: because it requires going back to
basic logical constructs every time, it does not scale.

Another example where we can contrast the two styles is any program
that computes the maximum of an array. In the atomic-assertion style,
the postcondition will be written

\begin{center}
\begin{adjustwidth}{-1cm}{-1cm}
\begin{center}
\begin{tabular}[h]{lp{4cm}}
  \ec{$\forall k \in \integers$: a.lower <= k <= a.upper $\;$implies$\;$ a[k] <= Result}    
  & (Every element between bounds has a value smaller than \ec{Result}) \\
  \ec{$\exists k \in \integers$: a.lower <= k <= a.upper land a[k] = Result}    
  & (Some element between bounds has the value \ec{Result}).
\end{tabular}
\end{center}
\end{adjustwidth}
\end{center}
This property is the definition of the maximum and hence needs to be
written somewhere. If we define a function ``$\max$'' to capture this
definition, the specification becomes simply
\begin{equation*}
  \ec{Result} = \max (a) .
\end{equation*}

The difference between the two styles becomes critical when we come to
the invariant of programs computing an array's maximum. Two different
algorithms appear in Section~\ref{sec:searching}. The first
(Section~\ref{sec:max-one-variable}) is the most straightforward; it
moves an index $i$ from \ec{a.lower + 1} to \ec{a.upper}, updating
\ec{Result} if the current value is higher than the current result
(initialized to the first element \ec{a [a.lower]}).
With a domain theory on arrays, the function $\max$ will be available
as well as a notion of slice, where the slice \ec{a [i..j]} for
integers $i$ and $j$ is the array consisting of elements of $a$ in the
range $[i,j]$. Then the invariant is simply
\begin{equation*}
  \ec{Result} = \max (\ec{a [a.lower..i]}) ,
\end{equation*}
which is ensured by initialization and, on exit when \ec{i = a.upper},
yields the postcondition \ec{Result = $\;\max(a)$} (based on the
domain-theory property that \ec{a [a.lower..a.upper] =
  a}). The atomic-assertion invariant would be a variation of the
expanded postcondition:
\begin{center}
\begin{tabular}[h]{lp{4cm}}
  \ec{$\forall k \in \integers$: a.lower <= k <= i $\;$implies$\;$ a[k] <= Result}    
  \\
  \ec{$\exists k \in \integers$: a.lower <= k <= i land a[k] = Result}.    
\end{tabular}
\end{center}

Consider now a different algorithm for the same problem
(Section~\ref{sec:max-two-variable}), which works by exploring the
array from both ends, moving the left cursor $i$ up if the element at
$i$ is less than the element at $j$ and otherwise moving the right
cursor $j$ down.  The atomic-assertion invariant can be written with
an additional level of quantification:
\begin{equation}
  \exists m: 
  \left(
    \begin{array}{rccc}
      \forall k \in \integers: & \ec{a.lower} \leq k \leq \ec{a.upper} & \ec{implies} & a[k] \leq m \\
      \exists k \in \integers: & i \leq k \leq j & \ec{and} & a[k] = m
    \end{array}
  \right) .
\label{eq:max-twovar-atomic-inv}
\end{equation}
Alternatively, we can avoid quantifier alternation using the
characterization based on the complement property that the maximal
element is \emph{not} outside the slice \ec{a[i..j]}:
\begin{equation}
  \forall k \in \integers: \quad \ec{a.lower} \leq k < i \ \lor\ j < k \leq \ec{a.upper} \ \ \limpl\ \ \ec{a[k]} \leq \ec{a[i]} \ \lor\ \ec{a[k]} \leq \ec{a[j]} .
\label{eq:max-twovar-atomic-inv-v2}
\end{equation}
The form without quantifier alternation is more amenable to automated
reasoning, but it has the disadvantage that it requires additional
ingenuity and is not a straightforward modification of the invariant
for the one-way version of the algorithm.  More significantly for this
paper's point of view, both formulations
\eqref{eq:max-twovar-atomic-inv}--\eqref{eq:max-twovar-atomic-inv-v2}
give an appearance of complexity even though the invariant is
conceptually very simple, capturing in a nutshell the essence of the
algorithm (as noted earlier, one of the applications of a good
invariant is that it enables us to understand the core idea behind a
loop):
\begin{equation}
  \max (a) = \max (a [i..j]) .
\label{eq:max-twovar-domaintheory}
\end{equation}
In words: the maximum of the entire array is to be found in the slice
that has not been explored yet. On exit, where $i = j$, we are left
with a one-element slice, whose value (this is a small theorem of the
corresponding domain theory) is its maximum and hence the maximum of
the whole array. The domain-theory invariant makes the algorithm and
its correctness immediately clear. 

The domain-theory approach means that, before any attempt to reason
about an algorithm, we should develop an appropriate model of the
underlying domain, by defining appropriate concepts such as greatest
common divisor for algorithms on integers and slices and maximum for
algorithms on arrays, establishing the relevant theorems (for example
that $x > y \limpl \gcd (x, y) = \gcd (x - y, y)$ and that $\max(a
[i..i]) = a [i]$). These concepts and theorems need only be developed
once for every application domain of interest, not anew for every
program over that domain. The programs can then use the corresponding
functions in their assertions, in particular in the loop invariants.

The domain-theory approach takes advantage of standard abstraction
mechanism of mathematics. Its only practical disadvantage, for
assertions embedded in a programming language, is that the functions
over a domain (such as $\gcd$) must come from some library and, if
themselves written in the programming language, must satisfy strict
limitations; in particular they must be ``pure'' functions defined
without any reference to imperative constructs. This issue only
matters, however, for the practical embedding of invariants in
programs; it is not relevant to the conceptual discussion of
invariants, independent of any implementation concerns, which is the
focus of this paper.

For the same reason, this paper does not explore---except for
Section~\ref{sec:boogie-assess}---the often delicate trade-off
between succinctness of expression and amenability to automated
reasoning.  For example, the invariant
\eqref{eq:max-twovar-atomic-inv} is concisely captured as
\eqref{eq:max-twovar-domaintheory} in domain-theory form even if it
uses quantifier alternation; the different formulation
\eqref{eq:max-twovar-atomic-inv-v2} is not readily expressible in
terms of slice and maximum functions, but it may be easier to handle
by automatic theorem provers since complexity grows with quantifier
alternation~\cite{Pap93}.  This paper's focus is on developing and
understanding the essence of algorithms through loop invariants
presented at the right level of abstraction, largely independent of
the requirements posed by automated reasoning.
Section~\ref{sec:boogie-assess}, however, demonstrates that the
domain-theory approach is still practically applicable. 

The remainder of this article relies, for each class of algorithms, on
the appropriate domain theory, whose components (functions and
theorems) are summarized at the beginning of the corresponding
section. We will make no further attempt at going back to the
atomic-assertion style; the examples above should suffice to show how
much simplicity is gained through this policy.

\section{Classifying invariants} \label{sec:classification}
Loop invariants and their constituent clauses can be classified along
two dimensions:
\begin{itemize}
\item   By   their   role    with   respect   to   the   postcondition
  (Section~\ref{sec:classify-role}), leading us to distinguish between
  ``essential'' and ``bounding'' invariant properties.
\item By the transformation technique that yields the invariant from
  the postcondition (Section~\ref{sec:classify-gen-technique}). Here
  we have techniques such as uncoupling and constant relaxation.
\end{itemize}

\subsection{Classification by role} \label{sec:classify-role}

In the typical loop strategy described in
Section~\ref{sec:constr-view}, it is essential that successive
iterations of the loop body remain in the convergence regions where
the generalized form of the postcondition is defined. The
corresponding conditions make up the \emph{bounding invariant}; the
clauses describing the generalized postcondition is the
\emph{essential invariant}. The bounding invariant for the greatest
common divisor algorithm consists of the clauses
\begin{align*}
& \ec{Result} > 0 \\
& x > 0 .
\end{align*}
The essential clause is
\begin{equation*}
\gcd (\ec{Result}, x) = \gcd (a, b) ,
\end{equation*}
yielding the postcondition when \ec{Result = x}.

For the one-way maximum program, the bounding invariant is
\begin{equation*}
\ec{a.lower} \leq i \leq \ec{a.upper}
\end{equation*}
and the essential invariant is
\begin{equation*}
\ec{Result} = \max \:(\ec{a [a.lower..i]}) ,
\end{equation*}
yielding the postcondition when \ec{i = a.upper}. Note that the
essential invariant would not be defined without the bounding
invariant, since the slice \ec{a[1..i]} would be undefined (if \ec{i >
  a.upper}) or would be empty and have no maximum (if \ec{i <
  a.lower}).

For the two-way maximum program, the bounding invariant is
\begin{equation*}
\ec{a.lower <= i <= j <= a.upper}
\end{equation*}
and the essential invariant is
\begin{equation*}
\max (a) = \max (a [i..j]) ,
\end{equation*}
yielding the postcondition when \ec{i = j}. Again, the essential
invariant would not be always defined without the bounding invariant.

The separation between bounding and essential invariants is often
straightforward as in these examples. In case of doubt, the following
observation will help distinguish. The functions involved in the
invariant (and often, those of the postcondition) are often partial;
for example:
\begin{itemize}
\item $\gcd (u,v)$ is only defined if $u$ and $v$ are both non-zero (and, since we consider natural integers only in the example, positive).
\item For an array $a$ and an integer $i$, \ec{a[i]} is only defined if \ec{i $\in$ [a.lower..a.upper]}, and the slice \ec{a[i..j]} is non-empty only if \ec{[i..j] $\subseteq$ [a.lower..a.upper]}.
\item $\max (a)$ is only defined if the array $a$ is not empty.
\end{itemize}
Since the essential clauses, obtained by postcondition generalization,
use \ifcsur\else\linebreak\fi\ec{$\gcd$ (Result, x)} and (in the array algorithms) array
elements and maxima, the invariants must include the bounding clauses
as well to ensure that these essential clauses are meaningful.
A similar pattern applies to most of the invariants studied later.

\subsection{Classification by generalization technique} \label{sec:classify-gen-technique}

The essential invariant is a mutation (often, a weakening) of the
loop's postcondition. The following mutation techniques are
particularly common:
 \begin{describe}{\emph{Backward reasoning:}}
\item[\emph{Constant relaxation:}] replace a constant $n$ (more generally, an
  expression which does not change during the execution of the
  algorithm) by a variable $i$, and use $i = n$ as part or all of the
  exit condition. 
\end{describe}
Constant relaxation is the technique used in the one-way array maximum
computation, where the constant is the upper bound of the array. The
invariant generalizes the postcondition ``\ec{Result} is the maximum
of the array up to \ec{a.lower}'', where \ec{a.lower} is a constant,
with ``\ec{Result} is the maximum up to \ec{i}''. This condition is trivial
to establish initially for a non-empty array (take \ec{i} to be
\ec{a.lower}), easy to extend to an incremented $i$ (take \ec{Result}
to be the greater of its previous value and \ec{a[i]}), and yields
the postcondition when \ec{i} reaches \ec{a.upper}. As we will see in
Section~\ref{sec:binary-search}, binary search differs from sequential
search by applying double constant relaxation, to both the lower and
upper bounds of the array.

\begin{describe}{\emph{Backward reasoning:}}
\item[\emph{Uncoupling:}] replace a variable $v$ (often \ec{Result}) by two
  (for example \ec{Result} and \ec{x}), using their equality as part
  or all of the exit condition. 
\end{describe}
Uncoupling is used in the greatest common divisor algorithm.

\begin{describe}{\emph{Backward reasoning:}}
\item[\emph{Term dropping:}] remove a subformula (typically a conjunct),
  which gives a straightforward weakening. 
\end{describe}
Term dropping is used in the partitioning algorithm
(Section~\ref{sec:quick-partitioning}).

\begin{describe}{\emph{Backward reasoning:}}
\item[\emph{Aging:}] replace a variable (more generally, an expression) by an
  expression that represents the value the variable had at previous
  iterations of the loop. 
\end{describe}
Aging typically accommodates ``off-by-one'' discrepancies between when
a variable is evaluated in the invariant and when it is updated in the
loop body.

\begin{describe}{\emph{Backward reasoning:}}
\item[\emph{Backward reasoning:}] compute the loop's postcondition from
  another assertion by backward substitution. 
\end{describe}
Backward reasoning can be useful for nested loops, where the inner
loop's postcondition can be derived from the outer loop's invariant.

\section{The invariants of important algorithms} \label{sec:catalog}

The following subsections include a presentation of several algorithms, their loop invariants,  and their connection with each algorithm's postcondition.
Table~\ref{tab:list-of-algorithms} lists the algorithms and their category.
For more details about variants of the algorithms and their implementation, we refer to standard textbooks on algorithms \cite{MehlhornS2008,CLRS09,Knu11}.

\begin{table}[!ht]
\centering
\small
\tbl{The algorithms presented in Section~\ref{sec:catalog}.\label{tab:list-of-algorithms}}{%
\begin{tabular}{lll}
\textsc{Algorithm} & \multicolumn{1}{c}{\textsc{Type}} & \multicolumn{1}{c}{\textsc{Section}} \\
\hline
Maximum search (one variable) & searching & \S~\ref{sec:max-one-variable} \\
Maximum search (two variable) & searching & \S~\ref{sec:max-two-variable} \\
Sequential search in unsorted array & searching & \S~\ref{sec:linear-search} \\
Binary search & searching & \S~\ref{sec:binary-search} \\
Integer division & arithmetic & \S~\ref{sec:integer-div} \\
Greatest common divisor (with division) & arithmetic & \S~\ref{sec:GCD-division} \\
Exponentiation (by squaring) & arithmetic & \S~\ref{sec:exp} \\
Long integer addition & arithmetic & \S~\ref{sec:longadd} \\
Quick sort's partitioning & sorting & \S~\ref{sec:quick-partitioning} \\
Selection sort & sorting & \S~\ref{sec:selection-sort} \\
Insertion sort & sorting & \S~\ref{sec:insertion-sort} \\
Bubble sort (basic) & sorting & \S~\ref{sec:bubble-sort-basic} \\
Bubble sort (improved) & sorting & \S~\ref{sec:bubble-sort-improved} \\
Comb sort & sorting & \S~\ref{sec:comb-sort} \\
Knapsack with integer weights & dynamic programming & \S~\ref{sec:knapsack} \\
Levenstein distance & dynamic programming & \S~\ref{sec:levenshtein} \\
Rotating calipers algorithm & computational geometry & \S~\ref{sec:rot-calipers} \\
List reversal &  data structures & \S~\ref{sec:reversal} \\
Binary search trees &  data structures & \S~\ref{sec:bst} \\
PageRank algorithm & fixpoint & \S~\ref{sec:pagerank} 
\end{tabular}}
\end{table}

\subsection{Array searching} \label{sec:searching} 

Many algorithmic problems can be phrased as \emph{search} over data
structures---from the simple arrays up to graphs and other
sophisticated representations.  This section illustrates some of the
basic algorithms operating on arrays.

\subsubsection{Maximum: one-variable loop} \label{sec:max-one-variable}

The following routine \ec{max_one_way} returns the maximum element of
an unsorted array \ec{a} of bounds \ec{a.lower} and \ec{a.upper}. The
maximum is only defined for a non-empty array, thus the
precondition \ec{a.count >= 1}. The postcondition can be written
\begin{equation*}
\ec{Result} = \max (a) .
\end{equation*}

Writing it in slice form, as \ec{Result = $\max$ (a
  [a.lower..a.upper])} yields the invariant by constant relaxation of
either of the bounds. We choose the second one, \ec{a.upper}, yielding
the essential invariant clause
\begin{equation*}
\ec{Result} = \max (\ec{a [a.lower..i]}) .
\end{equation*}
Figure~\ref{code:max-one-variable} shows the resulting implementation of the algorithm.

\begin{figure}[!hbt]
\begin{lstlisting}
max_one_way (a: ARRAY [T$\,$]): T
  require
     a.count >= 1  -- (*\textit{a.count}*) is the number of elements of the array
  local
     i: INTEGER
  do
     from
        i := a.lower ; Result := a [a.lower]
     invariant
        a.lower <= i <= a.upper
        Result = $\max$ (a [a.lower, i])
     until
        i = a.upper
     loop
        i := i + 1
        if Result < a [i] then Result := a [i] end
     variant
        a.upper - i + 1
     end
  ensure
     Result = $\max$ (a)
  end
\end{lstlisting}
\caption{Maximum: one-variable loop.}
\label{code:max-one-variable}
\end{figure}

Proving initiation is trivial. Consecution relies on the domain-theory
property that
\begin{equation*}
\max (\ec{a [1..i+1]}) = \max (\max (\ec{a [1..i]}), \ec{a [i + 1]}) .
\end{equation*}

\subsubsection{Maximum: two-variable loop} \label{sec:max-two-variable}

The one-way maximum algorithm results from arbitrarily choosing to
apply constant relaxation to either \ec{a.lower} or (as in the above
version) \ec{a.upper}. Guided by a symmetry concern, we may choose
double constant relaxation, yielding another maximum algorithm
\ec{max_two_way} which traverses the array from both ends. If \ec{i}
and \ec{j} are the two relaxing variables, the loop body either
increases \ec{i} or decreases \ec{j}. When \ec{i = j}, the loop has
processed all of \ec{a}, and hence \ec{i} and \ec{j} indicate the maximum
element.

The specification (precondition and postcondition) is the same as for
the previous algorithm.
Figure~\ref{code:max-two-variable} shows an implementation.

\begin{figure}[!htb]
\begin{lstlisting}
max_two_way (a: ARRAY [T$\,$]): T
  require
     a.count >= 1
  local
     i, j: INTEGER
  do
     from
        i := a.lower ;  j := a.upper
     invariant
        a.lower <= i <= j <= a.upper
        $\max$ (a [i..j]) = $\max$ (a)
     until
        i = j
     loop
        if a [i] > a [j] then j := j - 1 else i := i + 1 end
     variant
        j - i
     end
     Result := a [i]
  ensure
     Result = $\max$ (a)
  end
\end{lstlisting}
\caption{Maximum: two-variable loop.}
\label{code:max-two-variable}
\end{figure}

It is again trivial to prove initiation. Consecution relies on the following
two domain-theory properties:
\begin{align}
j > i \;\land\; \ec{a [i] >= a [j]} &\quad\limpl\quad \max (\ec{a [i..j]}) = \max (\ec{a [i..j - 1]}) \\
i < j \;\land\; \ec{a [j] >= a [i]} &\quad\limpl\quad \max (\ec{a [i..j]}) = \max (\ec{a [i + 1..j]}) .
\end{align}

\subsubsection{Search in an unsorted array} \label{sec:linear-search}

The following routine \ec{has_sequential} returns the position of an
occurrence of an element \ec{key} in an array \ec{a} or, if \ec{key}
does not appear, a special value. The algorithm applies to any
sequential structure but is shown here for arrays. For simplicity, we
assume that the lower bound \ec{a.lower} of the array is 1, so that we
can choose 0 as the special value. Obviously this assumption is easy
to remove for generality: just replace 0, as a possible value for
\ec{Result,} by \ec{a.lower - 1.}

The specification may use the domain-theory notation \ec{elements (a)}
to express the set of elements of an array \ec{a}. A simple form of
the postcondition is
\begin{equation}
\ec{Result /= 0} \quad\liff\quad \ec{key} \in \ec{elements(a)} ,
\label{eq:post-search-simple}
\end{equation}
which just records whether the \ec{key} has been found. We will
instead use a form that also records where the element appears if
present:
\begin{align}
\ec{Result /= 0} &\quad\limpl\quad  \ec{key = a [Result]}  \label{post:found} \\
\ec{Result = 0}  &\quad\limpl\quad  \ec{key} \not\in \ec{elements (a)} , \label{post:not-found}
\end{align}
to which we can for clarity prepend the bounding clause
\begin{equation*}
\ec{Result} \in \ec{[0..a.upper]}
\end{equation*}
to make it explicit that the array access in \eqref{post:found} is
defined when needed. 

\begin{figure}[!t]
\begin{lstlisting}
has_sequential (a: ARRAY [T$\,$]; key: T): INTEGER
  require
    a.lower = 1  -- For convenience only, may be removed (see text).
  local
    i: INTEGER
  do
    from 
      i := 0 ; Result := 0
    invariant
      0 <= i <= a.count
      Result $\in [0,i]$
      Result /= 0 $\ \limpl\ $  key = a [Result]
      Result = 0  $\ \limpl\ $  key $\not\in$ elements (a [1..i])
    until 
      i = a.upper
    loop
      i := i + 1
      if a [i] = key then Result := i end
    variant
      a.upper - i + 1
    end
  ensure
      Result $\in [0,a.upper]$
      Result /= 0 $\ \limpl\ $  key = a [Result]
      Result = 0  $\ \limpl\ $  key $\not\in$ elements (a)
end
\end{lstlisting}
  \caption{Search in an unsorted array.}
  \label{code:linear-search}
\end{figure}

If in \eqref{post:not-found} we replace \ec{a} by \ec{a [1..a.upper]},
we obtain the loop invariant of sequential search by constant
relaxation: introducing a variable \ec{i} to replace either of the
bounds \ec{1} and \ec{a.upper.} Choosing the latter yields the
following essential invariant:
\begin{equation*}
  \begin{array}{lcl}
    \ec{Result} \in [0,i] \\
    \ec{Result /= 0} &\limpl &  \ec{key = a [Result]}  \\
    \ec{Result = 0}  &\limpl &  \ec{key} \not\in \ec{elements (a [1..i])} ,
  \end{array}
\end{equation*}
leading to an algorithm that works on slices \ec{[1..i]} for increasing
\ec{i}, starting at 0 and with bounding invariant \ec{0 <= i <=
  a.count}, as shown in Figure~\ref{code:linear-search}.\footnote{Note
  that in this example it is OK for the array to be empty, so there is
  no precondition on \ecfootnote{a.upper,} although general properties of
  arrays imply that $\!$\ecfootnote{a.upper >= 0;}$\;$ the value 0 corresponds to an
  empty array.}

To avoid useless iterations the exit condition may be replaced by \ecplain{i = a.upper lor}\ec{Result > 0.}

To prove initiation, we note that initially \ec{Result} is 0 and the slice \ec{a [1..i]} is empty. Consecution follows from the domain-theory property that, for all \ec{1 <= i < a.upper}:
\begin{equation*}
  \ec{key} \in \ec{elements} (\ec{a [1..i+1]})  \ \liff\ 
  \ec{key} \in \ec{elements} (\ec{a [1..i]}) \lor \ec{key} = \ec{a [i+1]} .
\end{equation*}

\subsubsection{Binary search} \label{sec:binary-search}

Binary search works on sorted arrays by iteratively halving a segment
of the array where the searched element may occur. The search
terminates either when the element is found or when the segment
becomes empty, implying that the element appears nowhere in the array.

As already remarked by Knuth many years ago~\cite[Vol.~3, Sec.~6.2.1]{Knu11}:
\begin{quote}
  Although the basic idea of binary search is comparatively
  straightforward, the details can be surprisingly tricky, and many
  programmers have done it wrong the first few times they tried.
\end{quote}
Reasoning carefully on the specification (at the domain-theory level)
and the resulting invariant helps avoid mistakes.

For the present discussion it is interesting that the postcondition is
the same as for sequential search (Section~\ref{sec:linear-search}),
so that we can see where the generalization strategy differs, taking
advantage of the extra property that the array is sorted.

The algorithm and implementation now have the precondition
\[
\ec{sorted (a)} ,
\]
where the domain-theory predicate \ec{sorted (a),} defined as
\[
 \forall j \in \ec{[a.lower..a.upper - 1]}: \ec{a [j]} \leq \ec{a [j+1]} ,
\]
expresses that an array is sorted upwards. The domain theorem
on which binary search rests is that, for any value \ec{mid} in
\ec{[i..j]} (where \ec{i} and \ec{j} are valid indexes of the array), 
and any value \ec{key} of type \ec{T} (the type of the array elements):
\begin{equation}
\ec{key} \in \ec{elements} (\ec{a[i..j]}) \liff
\left(
\begin{array}{c}
\ec{key} \leq \ec{a[mid]} \land \ec{key} \in \ec{elements} (\ec{a[i..mid]}) \\
\lor \\
\ec{key} > \ec{a[mid]} \land \ec{key} \in \ec{elements} (\ec{a[mid+1..j]})
\end{array}
\right) .
\label{eq:domain-theory-binary-search}
\end{equation}
This property leads to the key insight behind binary search, whose
invariant follows from the postcondition by variable introduction, \ec{mid}
serving as that variable.

Formula~\eqref{eq:domain-theory-binary-search} is not symmetric with
respect to \ec{i} and \ec{j;} a symmetric version is possible, using
in the second disjunct, ``$\geq$'' rather than ``$>$'' and \ec{mid}
rather than \ec{mid + 1.} The form given in
\eqref{eq:domain-theory-binary-search} has the advantage of using two
mutually exclusive conditions in the comparison of \ec{key} to \ec{a
  [mid].} As a consequence, we can limit ourselves to a value \ec{mid}
chosen in \ec{[i..j - 1]} (rather than \ec{[i..j]}) since the first
disjunct does not involve \ec{j} and the second disjunct cannot hold
for \ec{mid = j} (the slice \ec{a [mid + 1..j]} being then empty). All
these observations and choices have direct consequences on the program
text, but are better handled at the specification (theory) level.

We will start for simplicity with the version
\eqref{eq:post-search-simple} of the postcondition that only records
presence or absence, repeated here for ease of reference:
\begin{equation}
\ec{Result /= 0} \quad\liff\quad \ec{key} \in \ec{elements(a)} .
\label{eq:post-search-simple-again}
\end{equation}
%
Duplicating the right-hand side of
\eqref{eq:post-search-simple-again}, writing \ec{a} in slice form
\ec{a[1..a.upper],} and applying constant relaxation twice, to the
lower bound 1 and the upper bound \ec{a.upper,} yields the essential
invariant:
\begin{equation}
\ec{key} \in \ec{elements(a[i..j])} \quad\liff\quad
\ec{key} \in \ec{elements(a)}
\label{eq:inv-binary-search}
\end{equation}
with the bounding invariant
\[
  \ec{1 <= i <= mid + 1 $\quad$land$\quad$ 1 <= mid <= j <= a.upper $\quad$land$\quad$ i <= j} \,,
\]
which combines the assumptions on \ec{mid} necessary to apply
\eqref{eq:domain-theory-binary-search}---also assumed in
\eqref{eq:inv-binary-search}---and the additional knowledge that
\ec{1 <= i} and \ec{j <= a.upper.} 

The attraction of this presentation is that:
\begin{itemize}
\item The two clauses \ec{key <= a[mid]} and \ec{key > a[mid]} of
  \eqref{eq:domain-theory-binary-search} are easy-to-test
  complementary conditions, suggesting a loop body that preserves the
  invariant by testing \ec{key} against \ec{a [mid]} and going left or
  right as a result of the test.
\item When \ec{i = j}---the case that serves as exit condition---the
  left side of the equivalence \eqref{eq:inv-binary-search} reduces
  to \ec{key = a [i];} evaluating this expression tells us whether
  \ec{key} appeared at all in the entire array, the information we
  seek. In addition, we can obtain the stronger postcondition,
  \eqref{post:found}--\eqref{post:not-found}, which
  gives \ec{Result} its precise value, by simply assigning \ec{i} to
  \ec{Result.}
\end{itemize}
This leads to the implementation in Figure~\ref{code:binary-search}.

\begin{figure}[!htb]
\begin{lstlisting}
has_binary (a: ARRAY [T$\,$]; key: T): INTEGER
  require
    a.lower = 1  -- For convenience, see comment about (*\textit{has\_sequential}*).
    a.count > 0
    sorted (a)
  local
    i, j, mid: INTEGER
  do
    from 
      i:= 1; j := a.upper; mid := 1; Result := 0
    invariant
      1 <= i <= mid + 1 land 1 <= mid <= j <= a.upper land i <= j
      key $\in$ elements (a[i..j]) $\ \liff\ $ key $\in$ elements (a)
    until 
      i = j
    loop
      mid := "(*A value in $[i..j-1]$*)"  -- In practice chosen as $i + (j - i) /\!/ 2$
      if a [mid] < key then i := mid +1 else j := mid end
    variant
      j - i 
    end
    if a [i] = key then Result := i end
    ensure
      0 <= Result <= n
      Result /= 0 $\ \limpl\ $  key = a [Result]
      Result = 0  $\ \limpl\ $  key $\not\in$ elements (a)
    end
\end{lstlisting}
  \caption{Binary search.}
  \label{code:binary-search}
\end{figure}

To prove initiation, we note that initially \ec{mid} is 1, 
so that \ec{mid $\:\in\:$ [i..j]} is true. Consecution follows
directly from \eqref{eq:domain-theory-binary-search}.




For the expression assigned to \ec{mid} in the loop, given in
pseudocode as ``A value in \ec{[i..j - 1]}'', the implementation indeed
chooses, for efficiency, the midpoint of the interval \ec{[i..j]},
which may be written \ec{i + (j - i) divOp 2} where ``\ec{divOp}''
denotes integer division. In an implementation, this form is to be
preferred to the simpler \ec{(i + j) divOp 2}, whose evaluation on a
computer may produce an integer overflow even when $i$, $j$, and their
midpoint are all correctly representable on the computer's number
system, but (because they are large) the sum $i + j$ is
not~\cite{Bloch-blog}. In such a case the evaluation of $j-i$ is
instead safe.

\subsection{Arithmetic algorithms} \label{sec:arithmetic}

Efficient implementations of the elementary arithmetic operations
known since grade school require non-trivial algorithmic skills and
feature interesting invariants, as the examples in this section
demonstrate.

\subsubsection{Integer division} \label{sec:integer-div}

The algorithm for integer division by successive differences
computes the integer quotient $q$ and the remainder $r$ of two
integers $m$ and $n$. The postcondition reads
\begin{align*}
&  0 \leq r < m  \\
&  n = m \cdot q + r .
\end{align*}

The loop invariant consists of a bounding clause and an essential
clause. The latter is simply an element of the postcondition:
\[
n = m \cdot q + r .
\]
The bounding clause weakens the other postcondition clause by keeping
only its first part:
\[
0 \leq r ,
\]
so that the dropped condition $r < m$ becomes the exit condition. As a
consequence, $r \geq m$ holds in the loop body, and the assignment
\ec{r := r - m} maintains the invariant property $0 \leq r$.
It is straightforward to prove the implementation in
Figure~\ref{code:integer-div} correct with respect to this
specification.

\begin{figure}[!ht]
\begin{lstlisting}
divided_diff (n, m: INTEGER): (q, r: INTEGER)
  require
    n >= 0
    m > 0
  do
    from
       r := n; q := 0
    invariant
       0 <= r
       n = m $\cdot$ q + r
    until
       r < m
    loop
       r := r - m
       q := q + 1
    variant r
    end
  ensure
    0 <= r < m
    n = m $\cdot$ q + r
  end
\end{lstlisting}
  \caption{Integer division.}
  \label{code:integer-div}
\end{figure}

\subsubsection{Greatest common divisor (with division)} \label{sec:GCD-division}

Euclid's algorithm for the greatest common divisor offers another
example where clearly separating between the underlying mathematical
theory and the implementation yields a concise and convincing
correctness argument. Sections~\ref{sec:basic-example} and
\ref{sec:domain-theory} previewed this example by using the form that
repeatedly subtracts one of the values from the other; here we will
use the version that uses division.

The greatest common divisor $\gcd (a, b)$ is the greatest integer that
divides both $a$ and $b$, defined by the following axioms, where $a$
and $b$ are nonnegative integers such that at least one of them is
positive (``\ec{modOp}'' denotes integer remainder):
\begin{align*}
& a \ec{modOp} \gcd (a, b) = 0 \\
& b \ec{modOp} \gcd (a, b) = 0 \\
& \forall d \in \naturals: (a \ec{modOp} d = 0) \land (b \ec{modOp} d = 0) \;\limpl\; d \leq \gcd (a, b) .
\end{align*}

From this definition follow several properties of the gcd function:
\begin{describe}{\emph{Commutativity:}}
\item[\emph{Commutativity:}] $\gcd(a,b) = \gcd(b,a)$

\item[\emph{Zero divisor:}] $\gcd(a,0) = a$

\item[\emph{Reduction:}] for $b > 0$, $\gcd(a,b) = \gcd(a \ec{modOp} b, b)$
\end{describe}
The following property of the remainder operation is also useful:
\begin{describe}{\emph{Commutativity:}}
\item[\emph{Nonnegativity:}] for integers $a \geq 0$ and $b >0$: $a \ec{modOp} b \geq 0$
\end{describe}

\begin{figure}[!t]
\begin{lstlisting}
gcd_Euclid_division (a, b: INTEGER): INTEGER
  require
    a > 0
    b >= 0
  local
    t, x, y: INTEGER
  do
    from
      x := a
      y := b
    invariant
      x > 0
      y >= 0
      gcd (x, y) = gcd (a, b)
    until 
      y = 0
    loop
      t := y
      y := x modOp y
      x := t
    variant y end
    Result := x
  ensure
    Result = $\gcd$ (a, b)
  end
\end{lstlisting}
  \caption{Greatest common divisor with division.}
  \label{code:GCD-division}
\end{figure}

From the obvious postcondition \ec{Result = $\:\gcd$ (a, b),} we obtain
the essential invariant in three steps:
\begin{enumerate}
\item By backward reasoning, derive the loop's postcondition $x =
  \gcd(a,b)$ from the routine's postcondition \ec{Result = $\:\gcd$
    (a,b).}
\item Using the \emph{zero divisor} property, rewrite it as $\gcd(x,0)
  = \gcd(a,b)$.
\item Apply constant relaxation, introducing variable $y$ to replace $0$.
\end{enumerate}
This gives the essential invariant $\gcd(x,y) = \gcd(a,b)$ together
with the bounding invariants $x > 0$ and $y \geq 0$. The
corresponding implementation is shown in
Figure~\ref{code:GCD-division}.\footnote{The variant is simply $y$,
  which is guaranteed to decrease at every iteration and can be bounded
  from below by the property $0 \leq x \ec{modOp} y < y$.}

Initiation is established trivially. Consecution follows from the
\emph{reduction} property. Note that, unlike in the difference version
(Section~\ref{sec:basic-example}), we can arbitrarily decide always to
divide $x$ by $y$, rather than having to find out which of the two
numbers is greater; hence the commutativity of $\gcd$ is not used in
this proof.

\subsubsection{Exponentiation by successive squaring} \label{sec:exp}

Suppose we do not have a built-in power operator and wish to compute
$m^n$. We may of course multiply $m$ by itself $n - 1$ times, but a more
efficient algorithm squares $m$ for all 1s values in the binary
representation of $n$. In practice, there is no need to compute this
binary representation. 

Given the postcondition
\[
\ec{Result} = m^n ,
\]
we first rewrite it into the obviously equivalent form $\ec{Result} \cdot 1^1 = m^n$.
Then, the invariant is obtained by double constant relaxation: the essential property 
\[
\ec{Result} \cdot x^y = m^n
\]
is easy to obtain initially (by setting \ec{Result,} $x$, and $y$ to
$1$, $m$, and $n$), yields the postcondition when $y = 0$, and can be
maintained while progressing towards this situation thanks to the
domain-theory properties
\begin{align}
x^{2z} &= (x^2)^{2z/2}  \label{eq:exp-square} \\
x^z   &= x \cdot x^{z-1} . \label{eq:exp-multiply}
\end{align}
Using only \eqref{eq:exp-multiply} would lead to the inefficient $(n -
1)$-multiplication algorithm, but we may use \eqref{eq:exp-square} for
even values of $y = 2z$.
This leads to the algorithm in Figure~\ref{code:exp}.

\begin{figure}[!t]
\begin{lstlisting}
power_binary (m, n: INTEGER): INTEGER
  require
    n >= 0
  local
    x, y: INTEGER
  do
    from
      Result := 1
      x := m
      y := n
    invariant
      y >= 0
      Result $\cdot\: x^y$ = $m^n$
    until  y = 0
    loop
      if y.is_even then
         x := x * x
         y := y divOp 2
      else
         Result := Result * x
         y := y - 1
      end
    variant y
    end
  ensure
    Result = $m^n$
  end
\end{lstlisting}
  \caption{Exponentiation by successive squaring.}
  \label{code:exp}
\end{figure}

Proving initiation is trivial. Consecution is a direct application of
the \eqref{eq:exp-square} and \eqref{eq:exp-multiply} properties.

\subsubsection{Long integer addition} \label{sec:longadd}

The algorithm for long integer addition computes the sum of two
integers $a$ and $b$ given in any base as arrays of positional digits
starting from the least significant position. For example, the array
sequence $\langle 3, 2, 0, 1 \rangle$ represents the number $138$ in
base 5 as $3\cdot 5^0 + 2\cdot 5^1 + 0\cdot 5^2 + 1 \cdot 5^3 = 138$.
For simplicity of representation, in this algorithm we use arrays
indexed by 0, so that we can readily express the value encoded in base
$b$ by an array \ec{a} as the sum:
\[
\sum_{k=0}^{\mathit{a.count}} a[k] \cdot b^k .
\]

The postcondition of the long integer addition algorithm has two
clauses. One specifies that the pairwise sum of elements in $a$ and
$b$ encodes the same number as \ec{Result}:
\begin{equation}
 \sum_{k = 0}^{n-1} (a[k] + b[k]) \cdot \ec{base}^k = 
 \sum_{k = 0}^{n} \ec{Result}[k] \cdot \ec{base}^k .
\label{eq:post-sum-longadd}
\end{equation}
\ec{Result} may have one more digit than $a$ or $b$; hence the
different bound in the two sums, where $n$ denotes $a$'s and $b$'s
length (normally written \ec{a.count} and \ec{b.count}).  The second
postcondition clause is the consistency constraint that \ec{Result} is
indeed a representation in base \ec{base}:
\begin{equation}
  \ec{has_base (Result, base)} ,
\label{eq:post-base-longadd}
\end{equation}
where the predicate \ec{has_base} is defined by a quantification over the array's length:
\[
  \ec{has_base (v, b)} \quad\liff\quad
  \forall k \in \naturals: 0 \leq k < \ec{v.count} \ \limpl\ 0 \leq v[k] < b .
\]

Both postcondition clauses appear mutated in the loop invariant.
First, we rewrite \ec{Result} in slice form \ec{Result [0..n]} in
\eqref{eq:post-sum-longadd} and \eqref{eq:post-base-longadd}.  The
first essential invariant clause follows by applying constant relaxation
to~\eqref{eq:post-base-longadd}, with the variable expression $i-1$
replacing constant~$n$:
\[
  \ec{has_base (Result [0..i - 1], base)} .
\]
The decrement is required because the loop updates $i$ at the
end of each iteration; it is a form of \emph{aging} (see
Section~\ref{sec:classify-gen-technique}).

To get the other part of the essential invariant, we first highlight the
last term in the summation on the right-hand side of
\eqref{eq:post-sum-longadd}:
\begin{equation*}
 \sum_{k = 0}^{n-1} (a[k] + b[k]) \cdot \ec{base}^k = 
 \ec{Result}[n]\cdot \ec{base}^n + \sum_{k = 0}^{n-1} \ec{Result}[k] \cdot \ec{base}^k .
\end{equation*}
We then introduce variables $i$ and \ec{carry}, replacing constants
\ec{n} and \ec{Result[n]}. Variable $i$ is the loop counter, also
mentioned in the other invariant clause; \ec{carry}, as the name
indicates, stores the remainder of each pairwise addition, which will
be carried over to the next digit.

The domain property that the integer division by $b$ of the sum of two
$b$-base digits $v_1, v_2$ is less than $b$ (all variables are integer):
\[
b > 0 \;\land\; v_1, v_2 \in [0..b - 1]
\quad\limpl\quad
(v_1 + v_2) \ec{divOp} b \in [0..b - 1]
\]
suggests the bounding invariant clause \ec{0 <= carry < base.}
Figure~\ref{code:longadd} shows the resulting implementation, where
the most significant digit is set after the loop before terminating.

\begin{figure}[!htb]
\begin{lstlisting}
addition (a, b: ARRAY [INTEGER]; 
          base: INTEGER): ARRAY [INTEGER]
  require
    base > 0
    a.count = b.count = n >= 1
    has_base (a, base)  -- (*\textit{a}*) is a valid encoding in base (*\textit{base}*)
    has_base (b, base)  -- (*\textit{b}*) is a valid encoding in base (*\textit{base}*)
    a.lower = b.lower = 0 -- For simplicity of representation
  local
    i, d, carry: INTEGER
  do
    Result := $\{0\}^{n+1}$  -- Initialize (*\textbf{Result}*) to an array of size $n + 1$ with all $0$s
    carry := 0
    from
      i := 0
    invariant
      $\sum_{k = 0}^{i-1}$(a[k] + b[k])$\cdot$base$^k$ = carry$\cdot$base$^i$ + $\sum_{k = 0}^{i-1}$Result[k]$\cdot$base$^k$
      has_base (Result [0..i-1], base)
      0 <= carry < base
    until
      i = n
    loop
      d := a [i] + b [i] + carry
      Result [i] := d modOp base
      carry := d divOp base
      i := i + 1
    variant n - i end
    Result [n] := carry
  ensure
    $\sum_{k = 0}^{n-1}$(a[k] + b[k])$\cdot$base$^k$ = $\sum_{k = 0}^{n}$Result[k]$\cdot$base$^k$
    has_base (Result, base)
  end
\end{lstlisting}
\caption{Long integer addition.}
\label{code:longadd}
\end{figure}

Initiation is trivial under the convention that an empty sum evaluates
to zero.  Consecution easily follows from the domain-theoretic
properties of the operations in the loop body, and in particular from
how the \ec{carry} and the current digit \ec{d} are set in each
iteration.

\subsection{Sorting} \label{sec:sorting}

A number of important algorithms sort an array based on pairwise
comparisons and swaps of  elements.The following domain-theory
notations will be useful for arrays $a$ and $b$:
\begin{itemize}
\item \ec{perm (a,b)} expresses that the arrays are permutations of
  each other (their elements are the same, each occurring the same
  number of times as in the other array).

\item \ec{sorted (a)} expresses that the array elements appear in
  increasing order: \ec{$\forall i \in$ [a.lower..a.upper - 1]: a [i] <= a [i
    + 1]}.
\end{itemize}

The sorting algorithms considered sort an array in place, with the
specification:
\begin{eifeq}
sort (a: ARRAY [T$\,$])
  require
    a.lower = 1
    a.count = n >= 1
  ensure
    perm (a, old a)
    sorted (a)
\end{eifeq}
The type \ec{T} indicates a generic type that is totally ordered and
provides the comparison operators $<$, $\leq$, $\geq$, and $>$.  The
precondition that the array be indexed from 1 and non-empty is a
simplification that can be easily dropped; we adopt it in this section
as it focuses the presentation of the algorithms on the interesting
cases. For brevity, we also use $n$ as an alias of \ec{a}'s length
\ec{a.count}.

The notation \ec{a[i..j] $\sim$ x}, for an array slice \ec{a [i..j]},
a scalar value \ec{x}, and a comparison operator $\sim$ among $<$,
$\leq$, $\geq$, and $>$, denotes that all elements in the slice
satisfy $\sim$ with respect to $x$: it is a shorthand for \ec{$\forall
  k \in [i..j]$: a[k] $\;\sim\;$ x.}

\subsubsection{Quick sort: partitioning} \label{sec:quick-partitioning}

At the core of the well-known quick sort algorithm lies the
partitioning procedure, which includes loops with an interesting
invariant; we analyze it in this section.

The procedure rearranges the elements in an array \ec{a} according to
an arbitrary value \ec{pivot} given as input: all elements in
positions up to \ec{Result} included are no larger than \ec{pivot},
and all elements in the other ``high'' portion (after position
\ec{Result}) of the array are no smaller than \ec{pivot.} Formally,
the postcondition is:
\begin{align*}
& \ec{0 <= Result <= n} \\
& \ec{perm (a, old a)} \\
& \ec{a [1..Result] <= pivot} \\
& \ec{a [Result + 1..n] >= pivot} \,.
\end{align*}
In the special case where all elements in \ec{a} are greater than or
equal to \ec{pivot}, \ec{Result} will be zero, corresponding to the
``low'' portion of the array being empty.

Quick sort works by partitioning an array, and then recursively
partitioning each of the two portions of the partition. The choice of
\ec{pivot} at every recursive call is crucial to guarantee a good
performance of quick sort. Its correctness, however, relies solely on
the correctness of \ec{partition}, not on the choice of \ec{pivot}.
Hence the focus of this section is on \ec{partition} alone.

The bulk of the loop invariant follows from the last three
clauses of the postcondition.  \ec{perm (a, old a)} appears unchanged
in the essential invariant, denoting the fact that the whole algorithm
does not change \ec{a}'s elements but only rearranges them.  The
clauses comparing \ec{a}'s slices to \ec{pivot} determine the rest of
the essential invariant, once we modify them by introducing loop variables
\ec{low} and \ec{high} decoupling and relaxing ``constant''
\ec{Result}:
\begin{align*}
& \ec{perm (a, old a)} \\
& \ec{a [1..low - 1] <= pivot} \\
& \ec{a [high + 1..n] >= pivot} \,.
\end{align*}
The formula \ec{low = high}---removed when decoupling---becomes the
main loop's exit condition. Finally, a similar variable introduction
applied twice to the postcondition \ec{0 <= Result <= n} suggests the
bounding invariant clause
\begin{align*}
& \ec{1 <= low <= high <= n} \,.
\end{align*}

\begin{figure}[!b]
\begin{lstlisting}
partition (a: ARRAY [T$\,$]; pivot: T): INTEGER
  require
    a.lower = 1
    a.count = n >= 1
  local
    low, high: INTEGER
  do
    from low := 1 ; high := n
    invariant
      1 <= low <= high <= n
      perm (a, old a)
      a [1..low - 1] <= pivot
      a [high + 1..n] >= pivot
    until low = high
    loop
      from -- This loop increases (*\textit{low}*)
      invariant -- Same as outer loop
      until low = high lor a[low] > pivot
      loop low := low + 1 end
      from -- This loop decreases (*\textit{high}*)
      invariant -- Same as outer loop
      until low = high lor a[high] < pivot
      loop high := high - 1 end
      a.swap (low, high) -- Swap the elements in positions (*\textit{low}*) and (*\textit{high}*)
    variant high - low end
    if a [low] >= pivot then
      low := low - 1
      high := low
    end
    Result := low
  ensure
    0 <= Result <= n
    perm (a, old a)
    a [1..Result] <= pivot
    a [Result + 1..n] >= pivot
  end
\end{lstlisting}
\caption{Quick sort: partitioning.}
\label{code:quick-partitioning}
\end{figure}

The slice comparison \ec{a [1..low - 1] <= pivot} also includes
aging of variable \ec{low}. This makes the invariant clauses fully
symmetric, and suggests a matching implementation with two inner loops
nested inside an overall outer loop.  The outer loop starts with
\ec{low = 1} and \ec{high = n} and terminates, with \ec{low = high},
when the whole array has been processed.  The first inner loop
increments \ec{low} until it points to an element that is larger than
\ec{pivot}, and hence is in the wrong portion of the array.
Symmetrically, the outer loop decrements \ec{high} until it points to
an element smaller than \ec{pivot}. After \ec{low} and \ec{high} are
set by the inner loops, the outer loop swaps the corresponding
elements, thus making progress towards partitioning the
array. Figure~\ref{code:quick-partitioning} shows the resulting
implementation. The closing conditional in the main routine's body
ensures that \ec{Result} points to an element no greater than
\ec{pivot}; this is not enforced by the loop, whose invariant leaves
the value of \ec{a [low]} unconstrained. In particular, in the special
case of all elements being no less than \ec{pivot}, \ec{low} and
\ec{Result} are set to zero after the loop.

In the correctness proof, it is useful to discuss the cases \ec{a
  [low] < pivot} and \ec{a [low] >= pivot} separately when proving
consecution.  In the former case, we combine \ec{a [1..low - 1] <=
  pivot} and \ec{a [low] < pivot} to establish the backward
substitution \ec{a [1..low] <= pivot}.  In the latter case, we combine
\ec{low = high}, \ec{a [high + 1..n] >= pivot} and \ec{a [low] >=
  pivot} to establish the backward substitution \ec{a [low..n] >=
  pivot}. The other details of the proof are straightforward.

\subsubsection{Selection sort} \label{sec:selection-sort}

Selection sort is a straightforward sorting algorithm based on a
simple idea: to sort an array, find the smallest element, put it in
the first position, and repeat recursively from the second position
on. Pre- and postcondition are the usual ones for sorting (see
Section~\ref{sec:sorting}), and hence require no further comment.

The first postcondition clause \ec{perm (a, old a)} is also an essential 
loop invariant: 
\begin{equation}
  \ec{perm (a, old a)} .
  \label{eq:perm-a}
\end{equation}
If we introduce a variable $i$ to iterate over the array, another
essential invariant clause is derived by writing \ec{a} in slice form
\ec{a[1..n]} and then by relaxing $n$ into $i$:
\begin{align}
& \ec{sorted (a [1..i])}  \label{eq:sorted-i}
\end{align}
with the bounding clause
\begin{align}
& 1 \leq i \leq n , \label{eq:i-in-bounds}
\end{align}
which ensures that the sorted slice \ec{a[1..i]} is always non-empty.
The final component of the invariant is also an essential weakening
of the postcondition, but is less straightforward to derive by
syntactic mutation. If we split \ec{a [1..n]} into the concatenation
\ec{a [1..i - 1] catOp a [i..n],} we notice that \ec{sorted (a [1..i -
  1] catOp a [i..n])} implies
\begin{align}
& \forall k \in [i..n]: \ec{a [1..i - 1] <= a[k]} \label{eq:smaller-left}
\end{align}
as a special case. Formula \eqref{eq:smaller-left} guarantees that the slice
\mbox{\ec{a[i..n]},} which has not been sorted yet, contains elements that are
no smaller than any of those in the sorted slice \ec{a[1..i - 1].}

The loop invariants \eqref{eq:perm-a}--\eqref{eq:i-in-bounds}
apply---possibly with minimal changes due to inessential details in
the implementation---for any sorting algorithm that sorts an array
sequentially, working its way from lower to upper indices.  To
implement the behavior specific to selection sort, we introduce an
inner loop that finds the minimum element in the slice \ec{a [i..n]},
which is not yet sorted.  To this end, it uses variables $j$ and $m$:
$j$ scans the slice sequentially starting from position $i + 1$; $m$
points to the minimum element found so far.  Correspondingly, the
inner loop's postcondition is \ec{a[m] <= a[i..n]}, which induces the
essential invariant clause
\begin{equation}
\ec{a [m]} \leq \ec{a [i..j - 1]}
\label{eq:ss-inner-lower}
\end{equation}
specific to the inner loop, by constant relaxation and aging.  The
outer loop's invariant \eqref{eq:smaller-left} clearly also applies to
the inner loop---which does not change $i$ or $n$---where it
implies that the element in position $m$ is an upper
bound on all elements already sorted:
\begin{equation}
\ec{a [1..i - 1]} \leq \ec{a [m]} .
\label{eq:ss-inner-upper}
\end{equation}
Also specific to the inner loop are more complex bounding invariants
relating the values of $i$, $j$, and $m$ to the array bounds:
\begin{align*}
& 1 \leq i < j \leq n + 1 \\
& i \leq m < j .
\end{align*}
The implementation in Figure~\ref{code:selection-sort} follows these
invariants.  The outer loop's only task is then to swap the
``minimum'' element pointed to by $m$ with the lowest available
position pointed to by $i$.

\begin{figure}[!htb]
\begin{lstlisting}
selection_sort (a: ARRAY [T$\,$])
  require
    a.lower = 1
    a.count = n >= 1
  local
    i, j, m: INTEGER
  do
    from i := 1
    invariant
      1 <= i <= n
      perm (a, old a)
      sorted (a [1..i])
      $\forall k \in [i..n]$: a [1..i - 1] <= a [k]
    until
      i = n
    loop
      from j := i + 1 ; m := i
      invariant
        1 <= i < j <= n + 1
        i <= m < j
        perm (a, old a)
        sorted (a [1..i])
        a [1..i - 1] <= a [m] <= a [i..j - 1]
      until
        j = n + 1
      loop
        if a [j] < a [m] then m := j end
        j := j + 1
      variant n - i - j end
      a.swap (i, m) (*\label{ln:selection-sort:swap}*) -- Swap the elements in positions (*\textit{i}*) and (*\textit{m}*)
      i := i + 1
    variant n - i end
  ensure
    perm (a, old a)
    sorted (a)
  end
\end{lstlisting}
\caption{Selection sort.}
\label{code:selection-sort}
\end{figure}

The most interesting aspect of the correctness proof is proving
consecution of the outer loop's invariant clause \eqref{eq:sorted-i},
and in particular that \ec{a[i] <= a[i + 1]}.  To this end, notice
that \eqref{eq:ss-inner-lower} guarantees that \ec{a [m]} is the
minimum of all elements in positions from $i$ to $n$; and
\eqref{eq:ss-inner-upper} that it is an upper bound on the other
elements in positions from $1$ to $i - 1$.  In particular,
\ec{a[m] <= a[i+1]} and \ec{a[i - 1] <= a[m]} hold before the swap
on line~\ref{ln:selection-sort:swap}. After the swap, \ec{a[i]} equals
the previous value of \ec{a[m]}, thus \ec{a[i - 1] <= a[i] <= a[i +
  1]} holds as required. A similar reasoning proves the inductiveness
of the main loop's other invariant clause \eqref{eq:smaller-left}.

\subsubsection{Insertion sort} \label{sec:insertion-sort}

Insertion sort is another sub-optimal sorting algorithm that is,
however, simple to present and implement, and reasonably efficient on
arrays of small size. As the name suggests, insertion sort hinges on
the idea of re-arranging elements in an array by inserting them in
their correct positions with respect to the sorting order; insertion
is done by shifting the elements to make room for insertion.  Pre- and
postcondition are the usual ones for sorting (see
Section~\ref{sec:sorting} and the comments in the previous
subsections).

The main loop's essential invariant is as in selection sort
(Section~\ref{sec:selection-sort}) and other similar algorithms, as it
merely expresses the property that the sorting has progressed up to
position $i$ and has not changed the array content:
\begin{align}
& \ec{sorted (a [1..i])}  \label{eq:ins-sorted-i} \\
& \ec{perm (a, old a)} . \label{eq:ins-perm-a}
\end{align}
This essential invariant goes together with the bounding clause $1
\leq i \leq n$.

The main loop includes an inner loop, whose invariant captures the
specific strategy of insertion sort. The outer loop's invariant
\eqref{eq:ins-perm-a} must be weakened, because the inner loop
overwrites \ec{a [i]} while progressively shifting to the right
elements in the slice \ec{a [1..j].} If a local variable \ec{v} stores
the value of \ec{a [i]} before entering the inner loop, we can weaken
\eqref{eq:ins-perm-a} as:
\begin{equation}
\ec{perm (a [1..j] catOp v catOp a[j + 2..n], old a)} , \label{eq:ins-perm-a-weaken}
\end{equation}
where ``\ec{catOp}'' is the concatenation operator; that is, \ec{a}'s
element at position $j+1$ is the current candidate for inserting
$v$---the value temporarily removed. After the inner loop terminates,
the outer loop will put $v$ back into the array at position $j+1$
(line~\ref{ln:restore-v} in Figure~\ref{code:insertion-sort}), thus
restoring the stronger invariant \eqref{eq:ins-perm-a} (and
establishing inductiveness for it).

The clause \eqref{eq:ins-sorted-i}, crucial for the
correctness argument, is also weakened in the inner loop. First, we
``age'' $i$ by replacing it with $i-1$; this corresponds to the fact
that the outer loop increments $i$ at the beginning, and will
then re-establish \eqref{eq:ins-sorted-i} only at the \emph{end} of
each iteration. Therefore, the inner loop can only assume the weaker
invariant:
\begin{equation}
\ec{sorted (a [1..i - 1])}  \label{eq:ins-sorted-i-weaken}
\end{equation}
that is not invalidated by shifting (which only temporarily duplicates
elements). Shifting has, however, another effect: since the slice
\ec{a[j + 1..i]} contains elements shifted up from the sorted portion,
the slice \ec{a[j + 1..i]} is itself sorted, thus the essential invariant:
\begin{equation}
\ec{sorted (a [j + 1..i])} .  \label{eq:ins-sorted-j-up-weaken}
\end{equation}
We can derive the pair of invariants
\eqref{eq:ins-sorted-i-weaken}--\eqref{eq:ins-sorted-j-up-weaken} from
the inner loop's postcondition \eqref{eq:ins-sorted-i}: write \ec{a
  [1..i]} as \ec{a [1..i - 1] catOp a[i..i]}; weaken the formula
\ec{sorted (a [1..i - 1] catOp a[i..i])} into the conjunction of
\ec{sorted( a [1..i - 1])} and \ec{sorted (a[i..i])}; replace one
occurrence of constant $i$ in the second conjunct by a fresh variable
$j$ and age to derive \ec{sorted (a [j + 1..i])}.

Finally, there is another essential invariant, specific to the inner
loop. Since the loop's goal is to find a position, pointed to by
$j+1$, before $i$ where $v$ can be inserted, its postcondition is:
\begin{equation}
\ec{v <= a [j + 1..i]} ,  \label{eq:ins-ordered-inner}
\end{equation}
which is also a suitable loop invariant, combined with a bounding
clause that constrains $j$ and $i$: 
\begin{equation}
\ec{0 <= j < i <= n} .  \label{eq:ins-bounding-inner}
\end{equation}
Overall, clauses
\eqref{eq:ins-perm-a-weaken}--\eqref{eq:ins-bounding-inner} are the
inner loop invariant; and Figure~\ref{code:insertion-sort} shows the
matching implementation.

\begin{figure}[!htb]
\begin{lstlisting}
insertion_sort (A: ARRAY [T$\,$])
  require
    a.lower = 1  ;  a.count = n >= 1
  local
    i, j: INTEGER ; v : T
  do
    from i := 1
    invariant
      1 <= i <= n
      perm (a, old a)
      sorted (a [1..i])
    until i = n
    loop
      i := i + 1
      v := a [i]
      from j := i - 1
      invariant
        0 <= j < i <= n
        perm (a [1..j] catOp v catOp a[j + 2..n], old a)
        sorted (a [1..i - 1])
        sorted (a [j + 1..i])
        v $\,$<=$\,$ a [j + 1..i]
      until j = 0 or a [j] <= v
      loop
        a [j + 1] := a [j]
        j := j - 1
      variant j - i end
      a [j + 1] := v (*\label{ln:restore-v}*)
    variant n - i end
  ensure
    perm (a, old a)
    sorted (a)
  end
\end{lstlisting}
  \caption{Insertion sort.}
  \label{code:insertion-sort}
\end{figure}

As usual for this kind of algorithms, the crux of the correctness
argument is proving that the outer loop's essential invariant is
inductive, based on the inner loop's. The formal proof uses the
following informal argument. Formulas \eqref{eq:ins-sorted-i-weaken} and
\eqref{eq:ins-ordered-inner} imply that inserting $v$ at $j+1$ does
not break the sortedness of the slice \ec{a [1..j + 1]}.  Furthermore,
\eqref{eq:ins-sorted-j-up-weaken} guarantees that the elements in the
``upper'' slice \ec{a [j + 1..i]} are also sorted with \ec{a [j] <=
  a[j + 1] <= a[j + 2].}  (The detailed argument would discuss the
cases $j = 0$, $0 < j < i - 1$, and $j = i - 1$.) In all, the whole
slice \ec{a [1..i]} is sorted, as required by \eqref{eq:ins-sorted-i}.

\subsubsection{Bubble sort (basic)} \label{sec:bubble-sort-basic}

As a sorting method, bubble sort is known not for its performance but
for its simplicity~\cite[Vol.~3, Sec.~5.2.2]{Knu11}.  It relies on the
notion of \emph{inversion}: a pair of elements that are not ordered,
that is, such that the first is greater than the second. The
straightforward observation that an array is sorted if and only if it
has no inversions suggests to sort an array by iteratively removing
all inversions. Let us present invariants that match such a high-level
strategy, deriving them from the postcondition (which is the same as
the other sorting algorithms of this section).

\begin{figure}[!t]
\begin{lstlisting}
bubble_sort_basic (a: ARRAY [T$\,$])
  require
    a.lower = 1  ;  a.count = n >= 1
  local
    swapped: BOOLEAN
    i: INTEGER
  do
    from swapped := True
    invariant
      perm (a, old a)
      lnot swapped limplies sorted (a)
    until lnot swapped
    loop
      swapped := False
      from i := 1
      invariant
        1 <= i <= n
        perm (a, old a)
        lnot swapped limplies sorted (a [1..i])
      until i = n
      loop
        if a [i] > a [i + 1] then
          a.swap (i, i + 1)-- Swap the elements in positions $i$ and $i+1$
          swapped := True
        end
        i := i + 1
      variant n - i end
    variant |inversions (a)| (*\label{ln:bubble-outer-variant}*)
    end
  ensure
    perm (a, old a)
    sorted (a)
  end
\end{lstlisting}
  \caption{Bubble sort (basic version).}
  \label{code:bubble-sort-basic}
\end{figure}

The postcondition \ec{perm (a, old a)} that $a$'s elements be not
changed is also an invariant of the two nested loops used in bubble
sort. The other postcondition \ec{sorted (a)} is instead weakened, but
in a way different than in other sorting algorithms seen before. We
introduce a Boolean flag \ec{swapped}, which records if there is
\emph{some} inversion that has been removed by swapping a pair of
elements. When \ec{swapped} is false after a complete scan of the
array \ec{a}, no inversions have been found, and hence \ec{a} is
sorted. Therefore, we use \ec{lnot swapped} as exit condition of the
main loop, and the weakened postcondition
\begin{equation}
\ec{lnot swapped limplies sorted (a)}  \label{eq:bubble-notswapped-inv}
\end{equation}
as its essential loop invariant.

The inner loop performs a scan of the input array that compares all
pairs of adjacent elements and swaps them when they are
inverted. Since the scan proceeds linearly from the first element to
the last one, we get an essential invariant for the inner loop by replacing
$n$ by $i$ in \eqref{eq:bubble-notswapped-inv} written in slice form:
\begin{equation}
\ec{lnot swapped limplies sorted (a [1..i])} .  \label{eq:bubble-notswapped-i-inv}
\end{equation}
The usual bounding invariant \ec{1 <= i <= n} and the outer loop's
invariant clause \ec{perm (a, old a)} complete the inner loop
invariant. 

The implementation is now straightforward to write as in
Figure~\ref{code:bubble-sort-basic}.  The inner loop, in particular,
sets \ec{swapped} to \ec{True} whenever it finds some inversion while
scanning. This signals that more scans are needed before the array is
certainly sorted.

Verifying the correctness of the annotated program in
Figure~\ref{code:bubble-sort-basic} is easy, because the essential
loop invariants \eqref{eq:bubble-notswapped-inv} and
\eqref{eq:bubble-notswapped-i-inv} are trivially true in all
iterations where \ec{swapped} is set to \mbox{\ec{True}.} On the other
hand, this style of specification makes the termination argument more
involved: the outer loop's variant (line~\ref{ln:bubble-outer-variant}
in Figure~\ref{code:bubble-sort-basic}) must explicitly refer to the
number of inversions left in \ec{a}, which are decreased by complete
executions of the inner loop.

\subsubsection{Bubble sort (improved)} \label{sec:bubble-sort-improved}

The inner loop in the basic version of bubble sort---presented in
Section~\ref{sec:bubble-sort-basic}---always performs a complete scan
of the $n$-element array \ec{a}. This is often redundant, because
swapping adjacent inverted elements guarantees that the largest
misplaced element is sorted after each iteration. Namely, the largest
element reaches the rightmost position after the first iteration, the
second-largest one reaches the penultimate position after the second
iteration, and so on. This section describes an implementation of bubble
sort that takes advantage of this observation to improve the running
time.

\begin{figure}[!b]
\begin{lstlisting}
bubble_sort_improved (a: ARRAY [T$\,$])
  require
    a.lower = 1  ;  a.count = n >= 1
  local
    i, j: INTEGER
  do
    from i := n
    invariant
      1 <= i <= n
      perm (a, old a)
      sorted (a [i..n])
      i < n $\;\limpl\;$ a[1..i] <= a[i + 1]
    until i = 1
    loop
      from j := 1
      invariant
        1 <= i <= n
        1 <= j <= i
        perm (a, old a)
        sorted (a [i..n])
        i < n $\;\limpl\;$ a[1..i] <= a[i + 1]
        a [1..j] <= a[j]
      until j = i
      loop
        if a [j] > a [j + 1] then a.swap (j, j + 1) end
        j := j + 1
      variant i - j
      end
      i := i - 1
    variant i
    end
  ensure
    perm (a, old a)
    sorted (a)
  end
\end{lstlisting}
  \caption{Bubble sort (improved version).}
  \label{code:bubble-sort-improved}
\end{figure}

The improved version still uses two nested loops. The outer loop's
essential invariant has two clauses:
\begin{equation}
\ec{sorted (a [i..n])}
\label{eq:bubble-high-sorted}
\end{equation}
is a weakening of the postcondition that encodes the knowledge that
the ``upper'' part of array \ec{a} is sorted, and 
\begin{equation}
\ec{i < n} \;\limpl\; \ec{a[1..i] <= a[i + 1]}
\label{eq:bubble-high-sorted-2}
\end{equation}
specifies that the elements in the unsorted slice \ec{a[1..i]} are no
larger than the first ``sorted'' element \ec{a[i + 1]}.  The
expression \ec{a[1..i] <= a[i + 1]} is a mutation (constant relaxation
and aging) of \ec{a[1..n] <= a[n]}, which is, in turn, a domain
property following from the postcondition.  Variable $i$ is now used
in the outer loop to mark the portion still to be sorted;
correspondingly, \eqref{eq:bubble-high-sorted-2} is well-defined only
when \ec{i < n}, and the bounding invariant clause \ec{1 <= i <= n} is
also part of the outer loop's specification.

Continuing with the same logic, the inner loop's postcondition:
\begin{equation}
\ec{a [1..i] <= a[i]}  \label{eq:bubble-inner-post}
\end{equation}
states that the largest element in the slice \ec{a [1..i]} has
been moved to the highest position.  Constant relaxation, replacing
$i$ (not changed by the inner loop) with a fresh variable $j$, yields
a new essential component of the inner loop's invariant:
\begin{equation}
\ec{a [1..j] <= a[j]} .  \label{eq:bubble-inner-progress}
\end{equation}
The outer loop's invariant and the bounding clause 
\ec{1 <= j <= i} complete the specification of the inner
loop. Figure~\ref{code:bubble-sort-improved} displays the
corresponding implementation.

The correctness proof follows standard strategies. In particular, the
inner loop's postcondition \eqref{eq:bubble-inner-post}---i.e., the
inner loop's invariant when $j = i$---implies \ec{a[i - 1] <= a[i]}
as a special case. This fact combines with the other clause
\eqref{eq:bubble-high-sorted-2} to establish the inductiveness of the
main loop's essential clause:
\[
\ec{sorted (a[i..n])} .
\]
Finally, proving termination is trivial for this program because each
loop has an associated iteration variable that is unconditionally
incremented or decremented.

\subsubsection{Comb sort} \label{sec:comb-sort}

In an attempt to improve performance in critical cases, comb sort
generalizes bubble sort based on the observation that small elements
initially stored in the right-most portion of an array require a large
number of iterations to be sorted. This happens because bubble sort
swaps adjacent elements; hence it takes $n$ scans of an array of size
$n$ just to bring the smallest element from the right-most $n$th
position to the first one, where it belongs. Comb sort adds the
flexibility of swapping non-adjacent elements, thus allowing for a faster
movement of small elements from right to left. A sequence of
non-adjacent equally-spaced elements also conveys the image of a
comb's teeth, hence the name ``comb sort''.

Let us make this intuition rigorous and generalize the loop
invariants, and the implementation, of the basic bubble sort algorithm
described in Section~\ref{sec:bubble-sort-basic}.  Comb sort is also
based on swapping elements, therefore the---now
well-known---invariant \ec{perm (a, old a)} also applies to its two
nested loops. To adapt the other loop invariant
\eqref{eq:bubble-notswapped-inv}, we need a generalization of the
predicate \ec{sorted} that fits the behavior of comb sort. Predicate
\ec{gap_sorted (a, d)}, defined as:
\[
\ec{gap_sorted} (a, d) \quad\liff\quad 
\forall k \in \ec{[a.lower..a.upper - d]}: \ec{a [k]} \leq \ec{a [k + d]}
\]
holds for arrays \ec{a} such that the subsequence of $d$-spaced
elements is sorted. Notice that, for $d = 1$, \ec{gap_sorted} reduces
to \ec{sorted}:
\[
\ec{gap_sorted (a, 1)} \;\liff\; \ec{sorted (a)} .
\]
This fact will be used to prove the postcondition from the loop
invariant upon termination.

With this new piece of domain theory, we can easily generalize the
essential and bounding invariants of
Figure~\ref{code:bubble-sort-basic} to comb sort. The outer loop
considers decreasing gaps; if variable \ec{gap} stores the current
value, the bounding invariant
\[
1 \leq \ec{gap} \leq n
\]
defines its variability range. Precisely, the main loop starts with
with \ec{gap = n} and terminates with \ec{gap = 1}, satisfying the
essential invariant:
\begin{equation}
\ec{lnot swapped limplies gap_sorted (a, gap)} .  \label{eq:comb-notswapped-inv}
\end{equation}
The correctness of comb sort does not depend on how \ec{gap} is
decreased, as long as it eventually reaches 1; if \ec{gap} is
initialized to 1, comb sort behaves exactly as bubble sort. In
practice, it is customary to divide \ec{gap} by some chosen parameter
$c$ at every iteration of the main loop.

\begin{figure}[!t]
\begin{lstlisting}
comb_sort (a: ARRAY [T$\,$])
  require
    a.lower = 1  ;  a.count = n >= 1
  local
    swapped: BOOLEAN
    i, gap: INTEGER
  do
    from swapped := True ; gap := n
    invariant
      1 <= gap <= n
      perm (a, old a)
      lnot swapped $\;$limplies$\;$ gap_sorted (a, gap)
    until
      lnot swapped and gap = 1
    loop
      gap := $\max$ (1, gap divOp c)
       -- $c > 1$ is a parameter whose value does not affect correctness
      swapped := False
      from i := 1
      invariant
        1 <= gap <= n
        1 <= i < i + gap <= n + 1
        perm (a, old a)
        lnot swapped $\;$limplies$\;$ gap_sorted (a [1..i - 1 + gap], gap)
      until 
        i + gap = n + 1
      loop
        if a [i] > a[i + gap] then
          a.swap (i, i + gap)
          swapped := True
        end
        i := i + 1
      variant n + 1 - gap - i end
    variant |inversions (a)| end
  ensure
    perm (a, old a)
    sorted (a)
  end
\end{lstlisting}
  \caption{Comb sort.}
  \label{code:comb-sort}
\end{figure}

Let us now consider the inner loop, which compares and swaps the
subsequence of $d$-spaced elements. The bubble sort invariant
\eqref{eq:bubble-notswapped-i-inv} generalizes to:
\begin{equation}
\ec{lnot swapped limplies gap_sorted (a [1..i - 1 + gap], gap)}  \label{eq:comb-notswapped-i-inv}
\end{equation}
and its matching bounding invariant is:
\[
1 \leq i < i + \ec{gap} \leq n +1
\]
so that when $i = n + 1 + \ec{gap}$ the inner loop terminates and
\eqref{eq:comb-notswapped-i-inv} is equivalent to
\eqref{eq:comb-notswapped-inv}. This invariant follows from constant
relaxation and aging; the substituted expression \ec{i - 1 + gap} is
more involved, to accommodate how $i$ is used and updated in the inner
loop, but is otherwise semantically straightforward.

The complete implementation is shown in
Figure~\ref{code:comb-sort}. The correctness argument is exactly as
for bubble sort in Section~\ref{sec:bubble-sort-basic}, but exploits
the properties of the generalized predicate \ec{gap_sorted} instead of
the simpler \ec{sorted}.

\subsection{Dynamic programming} \label{sec:dynamic}

Dynamic programming is an algorithmic technique used to compute
functions that have a natural recursive definition. Dynamic
programming algorithms construct solutions iteratively and store the
intermediate results, so that the solution to larger instances can
reuse the previously computed solutions for smaller instances. This
section presents a few examples of problems that lend themselves to
dynamic programming solutions.

\subsubsection{Unbounded knapsack problem with integer weights} \label{sec:knapsack}

We have an unlimited collection of items of $n$ different types. An
item of type $k$, for $k = 1, \ldots, n$, has weight $w[k]$ and value
$v[k]$. The unbounded knapsack problem asks what is the maximum
overall value that one can carry in a knapsack whose weight limit is a
given \ec{weight.} The attribute ``unbounded'' refers to the fact that
we can pick as many object of any type as we want: the only limit is
given by the input value of \ec{weight,} and by the constraint that
we cannot store fractions of an item---either we pick it or we
don't.

Any vector $s$ of $n$ nonnegative integers defines a selection of
items, whose \emph{overall weight} is given by the scalar product:
\begin{equation*}
s \cdot w \quad=\quad \sum_{1 \leq k \leq n} s[k]w[k]
\end{equation*}
and whose \emph{overall value} is similarly given by the scalar
product $s \cdot v$. Using this notation, we introduce the
domain-theoretical function \ec{max_knapsack} which defines the
solution of the knapsack problem given a weight limit $b$ and items of
$n$ types with weight and value given by the vectors $w$ and $v$:
\begin{equation*}
\ec{max_knapsack}\,(b, v, w, n) = \kappa
\liff
\left(
\begin{array}{rl}
\exists s \in \naturals^{n}: & s \cdot w \leq b \;\land\; s \cdot v = \kappa \\
 \land \\
\forall t \in \naturals^{n}: & t \cdot w \leq b \;\limpl\; t \cdot v \leq \kappa 
\end{array}
\right) ,
\end{equation*}
that is, the largest value achievable with the given limit.  Whenever
weights $w$, values $v$, and number $n$ of item types are clear from
the context, we will abbreviate \ec{max_knapsack (b, v, w, n)} by
just $K(b)$.

The unbounded knapsack problem is
NP-complete~\cite{GJ79,knapsack-book}.  It is, however, \emph{weakly}
NP-complete~\cite{Pap93}, and in fact it has a nice solution
with pseudo-poly\-no\-mi\-al complexity based on a recurrence
relation, which suggests a straightforward dynamic programming
algorithm.  The recurrence relation defines the value of $K(b)$ based
on the values $K(b')$ for $b' < b$.

The base case is for $b = 0$. If we assume, without loss of
generality, that no item has null weight, it is clear that we cannot
store anything in the knapsack without adding some weight, and hence
the maximum value attainable with a weight limit zero is also zero:
$K(0) = 0$.  Let now $b$ be a generic weight limit greater than
zero. To determine the value of $K(b)$, we make a series of attempts
as follows. First, we select some item of type $k$, such that $w[k]
\leq b$. Then, we recursively consider the best selection of items for
a weight limit of $b - w[k]$; and we set up a new selection by adding one
item of type $k$ to it. The new configuration has weight no greater
than $b - w[k] + w[k] = b$ and value
\[
v[k] + K(b - w[k]) ,
\]
which is, by inductive hypothesis, the largest achievable by adding an
element of type $k$. Correspondingly, the recurrence relation defines
$K(b$) as the maximum among all values achievable by adding an object
of some type:
\begin{equation}
K(b) =
\begin{cases}
0   &   b = 0 \\
\max \Big\{ v[k] + K(b - w[k]) \;\big\vert\; k \in [1..n] \text{ and } 0 \leq w[k] \leq b \Big\} 
& b > 0 .
\end{cases}
\label{eq:K-b-def}
\end{equation}

The dynamic programming solution to the knapsack problem presented in
this section computes the recursive definition \eqref{eq:K-b-def} for
increasing values of $b$. It inputs arrays $v$ and $w$ (storing the
values and weights of all elements), the number $n$ of element types,
and the weight bound \ec{weight}.  The precondition requires that
\emph{weight} be nonnegative, that all element weights $w$ be
positive, and that the arrays $v$ and $w$ be indexed from $1$ to $n$:
\begin{align*}
  & \ec{weight} \geq 0 \\
  & w > 0 \\
  & \ec{v.lower} = \ec{w.lower} = 1 \\
  & \ec{v.upper} = \ec{w.upper} = n \,.
\end{align*}
The last two clauses are merely for notational convenience and could
be dropped.  The postcondition states that the routine returns the
value $K(\ec{weight})$ or, with more precise notation:
\begin{equation*}
\ec{Result} = \ec{max_knapsack (weight, v, w, n)} .
\end{equation*}

The main loop starts with $b = 0$ and continues with unit increments
until $b = \ec{weight}$; each iteration stores the value of $K(b)$ in
the local array $m$, so that \ec{m [weight]} will contain the final
result upon termination. Correspondingly, the main loop's
essential invariant follows by constant relaxation:
\begin{itemize}
\item Variable \ec{m [weight]} replaces constant \ec{Result}, thus
  connecting \ec{m} to the returned result.
\item The range of variables \ec{[1..b]} replaces constant \ec{weight}, thus expressing
  the loop's incremental progress.
\end{itemize}

The loop invariant is thus:
\begin{equation}
\forall y \in [0..b]: \quad m [y] = \ec{max_knapsack} (y, v, w, n) ,
\label{eq:knapsack-main-progress-inv-quantifiers}
\end{equation}
which goes together with the bounding clause
\begin{equation*}
\ec{0 <= b <= weight}
\end{equation*}
that qualifies $b$'s variability domain. With a slight abuse of notation, we concisely write \eqref{eq:knapsack-main-progress-inv-quantifiers}, and similar expressions, as:
\begin{equation}
m [0..b] = \ec{max_knapsack} ([0..b], v, w, n) .
\label{eq:knapsack-main-progress-inv}
\end{equation}

The inner loop computes the maximum of definition \eqref{eq:K-b-def}
iteratively, for all element types $j$, where $1 \leq j \leq n$. To
derive its essential invariant, we first consider its postcondition
(similarly as the analysis of selection sort in
Section~\ref{sec:selection-sort}). Since the inner loop terminates at
the end of the outer loop's body, the inner's postcondition is the
outer's invariant \eqref{eq:knapsack-main-progress-inv}. Let us
rewrite it by highlighting the value $m[b]$ computed in the latest
iteration:
\begin{align}
m [0..b - 1] & \ =\ \ec{max_knapsack} \;([0..b - 1], v, w, n) \label{eq:knapsack-progress-minus} \\
m[b] & \ =\ \ec{best_value} \;(b, v, w, n, n) . \label{eq:knapsack-inner-post}
\end{align}
Function \ec{best_value} is part of the domain theory for knapsack,
and it expresses the ``best'' value that can be achieved given a
weight limit of $b$, $j \leq n$ element types, and assuming that the values
$K(b')$ for lower weight limits $b' < b$ are known:
\begin{multline*}
\ec{best_value}\,(b, v, w, j, n) = 
\max \Big\{
v[k] + K(b-w[k]) \;\big\vert\;
k \in [1..j] \text{ and } 0 \leq w[k] \leq b
\Big\} .
\end{multline*}
If we substitute variable $j$ for constant $n$ in
\eqref{eq:knapsack-inner-post}, expressing the fact that the inner
loop tries one element type at a time, we get the inner loop essential 
invariant:
\begin{align*}
m [0..b - 1] & \ =\ \ec{max_knapsack} \;([0..b - 1], v, w, n) \\
m[b] & \ =\ \ec{best_value} \;(b, v, w, j, m) .
\end{align*}
The obvious bounding invariants \ec{0 <= b <= weight} and \ec{0 <= j
  <= n} complete the inner loop's
specification. Figure~\ref{code:knapsack} shows the corresponding
implementation.

\begin{figure}[!htb]
\begin{lstlisting}
knapsack (v, w: ARRAY [INTEGER]; n, weight: INTEGER): INTEGER
  require
    weight >= 0
    w > 0
    v.lower = w.lower = 1
    v.upper = w.upper = n
  local
    b, j: INTEGER
    m: ARRAY [INTEGER]
  do
    from b := 0 ; m [0] := 0
    invariant
      0 <= b <= weight
      m [0..b] = max_knapsack ([0..b], v, w, n)
    until b = weight
    loop
      b := b + 1
      from j := 0 ; m [b] := m [b - 1]
      invariant
        0 <= b <= weight
        0 <= j <= n
        m [0..b - 1] = max_knapsack ([0..b - 1], v, w, n)
        m [b] = best_value (b, v, w, j, n)
      until j = n
      loop
        j := j + 1
        if w [j] <= b and m [b] < v [j] + m [b - w [j]] then
          m [b] := v [j] + m [b - w [j]]
        end
      variant n - j end
    variant weight - b end
    Result := m [weight]
  ensure
    Result = max_knapsack (weight, v, w, n)
  end
\end{lstlisting}
  \caption{Unbounded knapsack problem with integer weights.}
  \label{code:knapsack}
\end{figure}

The correctness proof reverses the construction we highlighted
following the usual patterns seen in this section. In particular,
notice that:
\begin{itemize}
\item When $j = n$ the inner loop terminates, thus establishing \eqref{eq:knapsack-progress-minus} and \eqref{eq:knapsack-inner-post}.

\item \eqref{eq:knapsack-progress-minus} and
  \eqref{eq:knapsack-inner-post} imply
  \eqref{eq:knapsack-main-progress-inv} because the recursive
  definition \eqref{eq:K-b-def} for some $b$ only depends on the
  previous values for $b' < b$, and \eqref{eq:knapsack-progress-minus}
  guarantees that \ec{m} stores those values.
\end{itemize}

\subsubsection{Levenshtein distance} \label{sec:levenshtein}

The \emph{Levenshtein distance} of two sequences $s$ and $t$ is the
minimum number of elementary edit operations (deletion, addition, or
substitution of one element in either sequence) necessary to turn $s$
into $t$. The distance has a natural recursive definition:
\begin{equation*}
\begin{footnotesize}
\ec{distance} (s, t) = 
\begin{cases}
  0   &   m = n = 0 \\
  m &   m > 0, n = 0 \\
  n &   n > 0, m = 0 \\
  \ec{distance} \left( \!\!\begin{array}{l}
      s [1..m-1], 
      t [1..n-1]
    \end{array} \!\!\right) &
    m > 0, n > 0, s[m] = t[n] \\
 1 + \min \left(\!\! \begin{array}{l}
   \ec{distance} (s [1..m-1], t), \\
   \ec{distance} (s, t [1..n-1]), \\
   \ec{distance} (s[1..m-1], t [1..n-1])
 \end{array} \!\!\right)
 & m > 0, n > 0, s[m] \neq t[n] ,
\end{cases}
\end{footnotesize}
\end{equation*}
where $m$ and $n$, respectively, denote $s$'s and $t$'s length (written
\ec{s.count} and \ec{t.count} when $s$ and $t$ are arrays).  The first
three cases of the definition are trivial and correspond to when $s$,
$t$, or both are empty: the only way to get a non-empty string from an
empty one is by adding all the former's elements. If $s$ and $t$ are
both non-empty and their last elements coincide, then the same number
of operations as for the shorter sequences $s[1..m-1]$ and $t[1..n-1]$
(which omit the last elements) are needed. Finally, if $s$'s and $t$'s
last elements differ, there are three options: (1) delete $s[m]$ and
then edit $s[1..m-1]$ into $t$; (2) edit $s$ into $t[1..n-1]$ and then
add $t[n]$ to the result; (3) substitute $t[n]$ for $s[m]$ and then
edit the rest $s[1..m-1]$ into $t[1..n-1]$. Whichever of the options
(1), (2), and (3) leads to the minimal number of edit operations is the
Levenshtein distance.

It is natural to use a dynamic programming algorithm to compute the
Levenshtein distance according to its recursive definition.  The
overall specification, implementation, and the corresponding proofs,
are along the same lines as the knapsack problem of
Section~\ref{sec:knapsack}; therefore, we briefly present only the
most important details. The postcondition is simply
\[
\ec{Result} = \ec{distance} \:(s, t) .
\]

The implementation incrementally builds a bidimensional matrix $d$ of
distances such that the element $d[i,j]$ stores the Levenshtein
distance of the sequences $s[1..i]$ and $t[1..j]$. Correspondingly,
there are two nested loops: the outer loop iterates over rows of $d$,
and the inner loop iterates over each column of $d$. Their essential
invariants express, through quantification, the partial progress
achieved after each iteration:
\begin{align*}
  \forall h \in [0..i - 1], \forall k \in [0..n]&: d [h, k] = \ec{distance} (s [1..h], t [1..k]) \\
  \forall h \in [0..i - 1], \forall k \in [0..j - 1]&: d [h, k] = \ec{distance} (s [1..h], t [1..k]) .
\end{align*}
The standard bounding invariants on the loop variables $i$ and $j$
complete the specification.

\begin{figure}[!hbt]
\begin{lstlisting}
Levenshtein_distance (s, t: ARRAY [T$\,$]): INTEGER
  require
    s.lower = t.lower = 1
    s.count = m
    t.count = n
  local
    i, j: INTEGER
    d: ARRAY [INTEGER, INTEGER]
  do
    d := $\{0\}^{m+1} \times \{0\}^{n+1}$
    across [1..m] as i loop d [i,0] := i end (*\label{ln:levenshtein-init-begin}*)
    across [1..n] as j loop d [0,j] := j end (*\label{ln:levenshtein-init-end}*)

    across [1..m] as i
    invariant
      1 <= i <= m + 1
      $\forall h \in [0..i - 1], \forall k \in [0..m]$: d [h, k] = distance (s [1..h], t [1..k])
    loop
      across [1..n] as j
      invariant
        1 <= i <= m + 1
        1 <= j <= n + 1
        $\forall h \in [0..i - 1], \forall k \in [0..j - 1]$: d [h, k] = distance (s [1..h], t [1..k])
      loop
        if s [i] = t [j] then  
           d [i,j] := d [i - 1, j - 1]
        else
           d [i,j] := 1 + $\min$ (d [i - 1, j - 1], d [i, j - 1], d [i - 1, j])
        end
      end
    end
    Result := d [m, n]
  ensure
    Result = distance (s, t)
  end
\end{lstlisting}
  \caption{Levenshtein distance.}
  \label{code:levenshtein}
\end{figure}

Figure~\ref{code:levenshtein} shows the implementation, which uses the
compact \ec{across} notation for loops, similar to ``for'' loops in
other languages. The syntax 
\begin{eifeq}
across [a..b] invariant I as k loop B end
\end{eifeq}
is simply a shorthand for:
\begin{eifeq}[18mm]
from k := a invariant I until k = b + 1 loop B ; k := k + 1 end
\end{eifeq}
For brevity, Figure~\ref{code:levenshtein} omits the obvious loop
invariants of the initialization loops at lines~\ref{ln:levenshtein-init-begin} and \ref{ln:levenshtein-init-end}.


\subsection{Computational geometry: Rotating calipers} 
\label{sec:geometry}
\label{sec:rot-calipers}

\begin{figure}[!hbt]
  \centering
  
\subfigure[Initial jaw configuration: antipodal pair $(p_1, p_5)$.]{
\begin{tikzpicture}[x=15pt,y=15pt]
\coordinate (p1) at (0,0);
\coordinate (p7) at (4,1);
\coordinate (p6) at (5,3);
\coordinate (p5) at (3,4.2);
\coordinate (p4) at (1.5,4.4);
\coordinate (p3) at (-1.5,2.4);
\coordinate (p2) at (-2,1.4);

\draw [very thick,draw=blue!50!black!80,fill=blue!50!black!20] (p1) -- (p2) -- (p3) -- (p4) -- (p5) -- (p6) -- (p7) -- cycle; 

\node at ($(p1)+(0.1,0.3)$) {\scriptsize $p_1$};
\node at ($(p7)+(-0.2,0.2)$) {\scriptsize $p_2$};
\node at ($(p6)+(-0.4,-0.1)$) {\scriptsize $p_3$};
\node at ($(p5)+(-0.1,-0.25)$) {\scriptsize $p_4$};
\node at ($(p4)+(0.1,-0.3)$) {\scriptsize $p_5$};
\node at ($(p3)+(0.35,-0.15)$) {\scriptsize $p_6$};
\node at ($(p2)+(0.45,0.1)$) {\scriptsize $p_7$};

\draw[ultra thick, color=red!80!black!80] ($(p1)+(-3,0)$) -- node [below] {\ec{jaw_a}} ++(8,0);
\draw[ultra thick, color=green!70!black!80] ($(p4)+(-5,0)$) -- node [above] {\ec{jaw_b}} ++(8,0);

\draw[very thick, color=black] ($(p1)+(2,0)$) arc (0:13.63:2);
\draw[very thick, color=black] ($(p4)+(-2,0)$) arc (180:214:2);

\node at ($(p1)+(3,0)+(30:0.8)$) {\ec{angle_a}};
\node at ($(p4)+(-5.2,0)+(-20:2)$) {\ec{angle_b}};
\end{tikzpicture}
\label{fig:calipers-before}
}
\qquad\qquad
\subfigure[Jaws after one iteration: antipodal pair $(p_2, p_5)$.]{
\begin{tikzpicture}[x=15pt,y=15pt]
\coordinate (p1) at (0,0);
\coordinate (p7) at (4,1);
\coordinate (p6) at (5,3);
\coordinate (p5) at (3,4.2);
\coordinate (p4) at (1.5,4.4);
\coordinate (p3) at (-1.5,2.4);
\coordinate (p2) at (-2,1.4);

\draw [very thick,draw=blue!50!black!80,fill=blue!50!black!20] (p1) -- (p2) -- (p3) -- (p4) -- (p5) -- (p6) -- (p7) -- cycle; 

\node at ($(p1)+(0.1,0.3)$) {\scriptsize $p_1$};
\node at ($(p7)+(-0.2,0.2)$) {\scriptsize $p_2$};
\node at ($(p6)+(-0.4,-0.1)$) {\scriptsize $p_3$};
\node at ($(p5)+(-0.1,-0.25)$) {\scriptsize $p_4$};
\node at ($(p4)+(0.1,-0.3)$) {\scriptsize $p_5$};
\node at ($(p3)+(0.35,-0.15)$) {\scriptsize $p_6$};
\node at ($(p2)+(0.45,0.1)$) {\scriptsize $p_7$};

\fill[draw=black!25,top color=white, bottom color=black!25] (p1) -- ($(p1)+(2,0)$) arc (0:13.63:2) -- cycle;
\draw[ultra thick, color=red!80!black!80] ($(p1)+(13.63:6)$) -- node [below] {\ec{jaw_a}} ($(p1)+(193.63:3)$);
\draw[very thick, ->, color=black] ($(p1)+(2,0)$) arc (0:13.63:2);

\draw[draw=black!25,bottom color=white, top color=black!25] (p4) -- ($(p4)+(-2,0)$) arc (180:193.63:2) -- cycle;
\draw[ultra thick, color=green!70!black!80] ($(p4)+(193.63:5)$) -- node [above] {\ec{jaw_b}} ($(p4)+(13.63:3)$);
\draw[very thick, ->, color=black] ($(p4)+(-2,0)$) arc (180:193.63:2);

\end{tikzpicture}
\label{fig:calipers-after}
}

 \caption{The rotating calipers algorithm illustrated.}
  \label{fig:rot-calipers}
\end{figure}
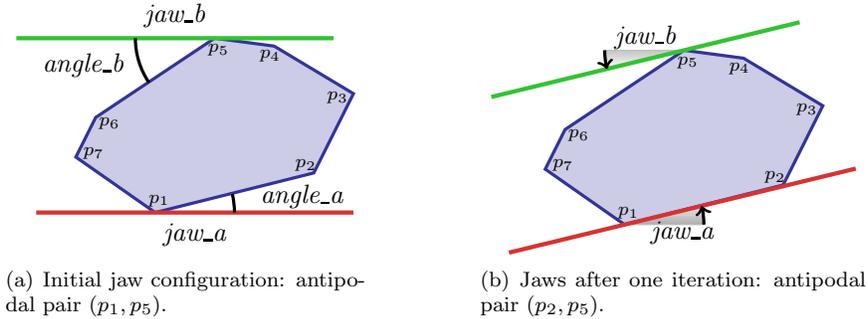

The \emph{diameter} of a polygon is its maximum width, that is the
maximum distance between any pair of its points. For a convex polygon,
it is clear that a pair of vertices determine the diameter (such as vertices $p_3$ and $p_7$ in Figure~\ref{fig:rot-calipers}). Shamos
showed~\citeyear{Shamos-thesis} that it is not necessary to check all $O(n^2)$ pairs
of vertices: his algorithm, described later, runs in time $O(n)$. The
correctness of the algorithm rests on the notions of \emph{lines of
  support} and antipodal points. A line of support is analogue to a
tangent: a line of support of a convex polygon $p$ is a line that
intersects $p$ such that the interior of $p$ lies entirely to one
side of the line. An antipodal pair is then any pair of $p$'s
vertices that lie on two \emph{parallel} lines of support. It is a
geometric property that an antipodal pair determines the diameter of
any polygon $p$, and that a convex polygon with $n$ vertices has $O(n)$
antipodal pairs.
Figure~\ref{fig:calipers-before}, for example, shows two parallel lines of support that identify the antipodal pair $(p_1, p_5)$.

Shamos's algorithms efficiently enumerates all antipodal pairs by
rotating two lines of support while maintaining them parallel. After a
complete rotation around the polygon, they have touched all antipodal
pairs, and hence the algorithm can terminate. Observing that two
parallel support lines resemble the two jaws of a caliper,
Toussaint~\citeyear{Tous83} suggested the name ``rotating calipers'' to
describe Shamos's technique.

\begin{figure}[!htb]
\begin{lstlisting}
diameter_calipers (p: LIST [POINT$\,$]): INTEGER
  require
    p.count >= 3 ; p.is_convex (*\label{ln:calipers-pre}*)
  local
    jaw_a, jaw_b: VECTOR  -- Jaws of the caliper
    n_a, n_b: NATURAL  -- Pointers to vertices of the polygon
    angle_a, angle_b: REAL  -- Angle measures
  do
    n_a := (*``Index, in $p$, of the vertex with the \emph{minimum} $y$ coordinate''*)
    n_b := (*``Index, in $p$, of the vertex with the \emph{maximum} $y$ coordinate''*)

    jaw_a := (*``Horizontal direction from*) p[n_a] (*pointing towards negative''*)
    jaw_b := (*``Horizontal direction from*) p[n_b] (*pointing towards positive''*)
    from
      total_rotation := 0
      Result := |p[n_a] - p[n_b]| -- Distance between pair of vertices
    invariant
      parallel (jaw_a, jaw_b) -- Jaws are parallel (*\label{ln:calipers-inv-parallel}*)
      0 < Result <= diameter (p) -- (*\textbf{Result}*) increases until (*\textit{diameter}$(p)$*) (*\label{ln:calipers-inv-result}*)
      0 <= total_rotation < 360 (*\label{ln:calipers-inv-rotation}*)
    until total_rotation >= 180  -- All antipodal pairs considered
    loop
      angle_a := (*``Angle between*) p[n_a] (*and the next vertex in*) p (*''*)
      angle_b := (*``Angle between*) p[n_b] (*and the next vertex in*) p (*''*)
      if angle_a < angle_b then
        -- Rotate (*\textit{jaw\_a}*) to coincide with the edge (*$\mathit{p[n\_a]}$---$\mathit{p[n\_a].next}$*)
        jaw_a := jaw_a + angle_a
        -- Rotate (*\textit{jaw\_b}*) by the same amount
        jaw_b := jaw_b + angle_a
        -- Next current point (*\textit{n\_a}*)
        n_a := (*``Index of vertex following $p[n\_a]$''*)
        -- Update total rotation
        total_rotation := total_rotation + angle_a
      else
        -- As in the (*\textbf{then}*) branch with $a$'s and $b$'s roles reversed
      end
      -- Update maximum distance between antipodal points
      Result := $\max$ (|p[n_a] - p[n_b]|, Result)
    variant 180 - total_rotation end
  ensure
    Result = diameter (p) (*\label{ln:calipers-post}*)
  end
\end{lstlisting}
  \caption{Diameter of a polygon with rotating calipers.}
  \label{code:rot-calipers}
\end{figure}

A presentation of the rotating calipers algorithm and a proof of its
correctness with the same level of detail as the algorithms in the
previous sections would require the development of a complex domain
theory for geometric entities, and of the corresponding implementation
primitives.  Such a level of detail is beyond the scope of this paper;
instead, we outline the essential traits of the specification and give
a description of the algorithm in pseudo-code in
Figure~\ref{code:rot-calipers}.\footnote{The algorithm in
  Figure~\ref{code:rot-calipers} is slightly simplified, as it does
  not deal explicitly with the special case where a line initially
  coincides with an edge: then, the minimum angle is zero, and hence
  the next vertex not on the line should be considered. This problem
  can be avoided by adjusting the initialization to avoid that a line
  coincides with an edge.}

The algorithm inputs a list $p$ of at least three points such that it
represents a convex polygon (precondition on
line~\ref{ln:calipers-pre}) and returns the value of the polygon's
diameter (postcondition on line~\ref{ln:calipers-post}).  

It starts by adjusting the two parallel support lines on the two
vertices with the maximum difference in $y$ coordinate, such as $p_1$
and $p_5$ in Figure~\ref{fig:calipers-before}. Then, it enters a loop
that computes the angle between each line and the next vertex on the
polygon, and rotates both jaws by the minimum of such angles. At each
iteration, it compares the distance between the new antipodal pair and
the currently stored maximum (in \ec{Result}), and updates the latter
if necessary. In the example of Figure~\ref{fig:calipers-before},
\ec{jaw_a} determines the smallest angle, and hence both jaws are
rotated by \ec{angle_a} in Figure~\ref{fig:calipers-after}.  Such an
informal description suggests the obvious bounding invariants that the
two jaws are maintained parallel
(line~\ref{ln:calipers-inv-parallel}), \ec{Result} varies between some
initial value and the final value \ec{diameter (p)}
(line~\ref{ln:calipers-inv-result}), and the total rotation of the
calipers is between zero and $180+180$
(line~\ref{ln:calipers-inv-rotation}). The essential invariant is,
however, harder to state formally, because it involves a subset of the
antipodal pairs reached by the calipers. A semi-formal presentation
is:
\[
\ec{Result} = \max 
\left\{
|p_1, p_2| \bigg| \begin{array}{l}
  p_1, p_2 \in p \quad\land \\
  \ec{reached ($p_1$, total_rotation)} \quad\land \\
  \ec{reached ($p_2$, total_rotation)}
\end{array}
\right\}
\]
whose intended meaning is that \ec{Result} stores the maximum distance
between all points $p_1, p_2$ among $p$'s vertices that can be reached
with a rotation of up to \ec{total_rotation} degrees from the initial
calipers' horizontal positions. 

\subsection{Algorithms on data structures}
\label{sec:datastructure}

Many data structures are designed around specific operations, which
can be performed efficiently by virtue of the characterizing
properties of the structures.  This section presents linked lists and
binary search trees and algorithms for some of such operations.  The
presentation of their invariants clarifies the connection between
data-structure properties and the algorithms' correct design.

\subsubsection{List reversal}
\label{sec:reversal}

Consider a list of elements of generic type \ec{G} implemented as a linked
list: each element's attribute \ec{next} stores a reference to the
next element in the list; and the last elements's \ec{next} attribute
is \ec{Void}. This section discusses the classic algorithm that
reverses a linked list iteratively.

We introduce a specification that abstracts some implementation
details by means of a suitable domain theory.  If \ec{list} is a
variable of type \ec{LIST [G]}---that is, a reference to the first
element---we lift the semantics of \ec{list} in assertions to denote
the \emph{sequence} of elements found by following the chain of
references until \ec{Void}. This interpretation defines finite
sequences only if \ec{list}'s reference sequence is acyclic, which we
write \ec{acyclic (list)}. Thus, the precondition of routine
\ec{reverse} is simply
\[
\ec{acyclic (list)} ,
\]
where \ec{list} is the input linked list.

For a sequence $s = s_1\,s_2\,\ldots\,s_n$ of elements of length $n
\geq 0$, its reversal \ec{rev (s)} is inductively defined as:
\begin{equation}
\ec{rev} (s) = 
\begin{cases}
\epsilon  &  n = 0 \\
\ec{rev} (s_2\,\ldots\,s_n) \:\ec{catOp}\: s_1 & n > 0 \,,
\end{cases}
\label{eq:rev-definition}
\end{equation}
where $\epsilon$ denotes the empty sequence and ``\ec{catOp}'' is the
concatenation operator. With this notation, \ec{reverse}'s postcondition is:
\[
\ec{list = rev (old list)}
\]
with the matching property that \ec{list} is still acyclic.

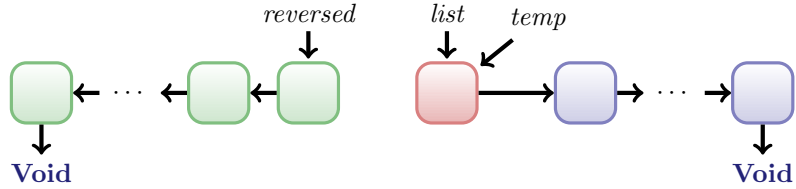
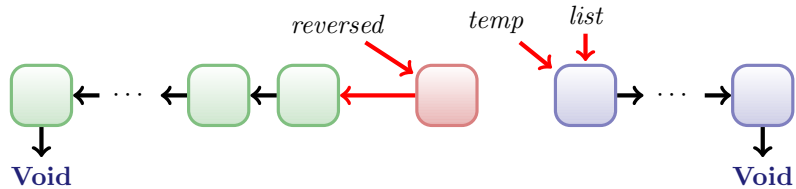
\begin{figure}[!htb]
  \centering
  \subfigure[Before executing the loop body.]{
  \begin{tikzpicture}[
  item/.style={rectangle, minimum size=8mm,very thick,rounded corners=2mm,draw=blue!50!black!50,font=\scriptsize, top color=white, bottom color=blue!50!black!20},
  revitem/.style={rectangle, minimum size=8mm,very thick,rounded corners=2mm,draw=green!50!black!50,font=\scriptsize, top color=white, bottom color=green!50!black!20},
  tomove/.style={rectangle, minimum size=8mm,very thick,rounded corners=2mm,draw=red!70!black!50,font=\scriptsize, top color=white, bottom color=red!70!black!30},
  node distance=3.5mm,
  pin distance=4mm,
  every pin edge/.style={<-, shorten <=1pt, ultra thick,black},
  ]

  \node (list) [tomove,pin=above:\ec{list},pin=above right:\ec{temp}] {};
  \node (i1) [item,right=10mm of list] {};
  \node (i3) [right=of i1,minimum size=8mm] {$\cdots$};
  \node (i4) [item,right=of i3] {};
  \node (iend) [below=4mm of i4,minimum size=4mm] {\ec{Void}};

  \node (rev) [revitem,left=1cm of list, pin=above:\ec{reversed}] {};
  \node (ri1) [revitem,left=of rev] {};
  \node (ri2) [left=of ri1,minimum size=8mm] {$\cdots$};
  \node (ri4) [revitem,left=of ri2] {};
  \node (riend) [below=4mm of ri4,minimum size=4mm] {\ec{Void}};

  \begin{scope}[->,black,ultra thick]
    \foreach \na / \nb in {list/i1,i1/i3,i3/i4,i4/iend,rev/ri1,ri1/ri2,ri2/ri4,ri4/riend}
    {
      \path (\na) edge (\nb);
    }
  \end{scope}
  \end{tikzpicture}
  \label{fig:reversal-before}
} \\

  \subfigure[After executing the loop body: new links are in red.]{
    \begin{tikzpicture}[
  item/.style={rectangle, minimum size=8mm,very thick,rounded corners=2mm,draw=blue!50!black!50,font=\scriptsize, top color=white, bottom color=blue!50!black!20},
  revitem/.style={rectangle, minimum size=8mm,very thick,rounded corners=2mm,draw=green!50!black!50,font=\scriptsize, top color=white, bottom color=green!50!black!20},
  tomove/.style={rectangle, minimum size=8mm,very thick,rounded corners=2mm,draw=red!70!black!50,font=\scriptsize, top color=white, bottom color=red!70!black!30},
  node distance=3.5mm,
  pin distance=4mm,
  every pin edge/.style={<-, shorten <=1pt, ultra thick,red},
  ]

  \node (list) [tomove,pin=above left:\ec{reversed}] {};
  \node (Plist) [above=5mm of list] {};
  \node (i1) [item,right=10mm of list,pin=above left:\ec{temp},pin=above:\ec{list}] {};
  \node (Ptemp) [above=4mm of i1] {};
  \node (i3) [right=of i1,minimum size=8mm] {$\cdots$};
  \node (i4) [item,right=of i3] {};
  \node (iend) [below=4mm of i4,minimum size=4mm] {\ec{Void}};

  \node (rev) [revitem,left=1cm of list] {};
  \node (Prev) [above=5mm of rev] {};
  \node (ri1) [revitem,left=of rev] {};
  \node (ri2) [left=of ri1,minimum size=8mm] {$\cdots$};
  \node (ri4) [revitem,left=of ri2] {};
  \node (riend) [below=4mm of ri4,minimum size=4mm] {\ec{Void}};

  \begin{scope}[->,black,ultra thick]
    \foreach \na / \nb in {i1/i3,i3/i4,i4/iend,rev/ri1,ri1/ri2,ri2/ri4,ri4/riend}
    {
      \path (\na) edge (\nb);
    }
  \end{scope}
  \begin{scope}[red,ultra thick,->]
    \path (list) edge (rev);
  \end{scope}
  \end{tikzpicture}
  \label{fig:reversal-after}
}

\caption{One iteration of the loop in routine \ec{reverse}.}
\label{fig:reversal}
\end{figure}

The domain theory for lists makes it possible to derive the loop
invariant with the usual techniques. Let us introduce a local variable
\ec{reversed}, which will store the iteratively constructed reversed
list. More precisely, every iteration of the loop removes one element
from the beginning of \ec{list} and moves it to the beginning of
\ec{reversed}, until all elements have been moved.  When the loop
terminates:
\begin{itemize}
\item \ec{list} points to an empty list;
\item \ec{reversed} points to the reversal of \ec{old list}.
\end{itemize}
Therefore, the routine concludes by assigning \ec{reversed} to
overwrite \ec{list}. Backward substitution yields the loop's
postcondition from the routine's:
\begin{equation}
\ec{reversed = rev (old list)} .
\label{eq:reversed-old-list}
\end{equation}
Using~\eqref{eq:rev-definition} for empty lists, we can equivalently
write \eqref{eq:reversed-old-list} as:
\begin{equation}
\ec{rev(list) catOp reversed = rev (old list)} ,
\label{eq:reversed-old-list-cat}
\end{equation}
which is the essential loop invariant, whereas \ec{list = Void} is the
exit condition. The other component of the loop invariant is the
constraint that \ec{list} and \ec{reversed} be
acyclic, also by mutation of the postcondition. 

\begin{figure}[!htb]
\begin{lstlisting}
reverse (list: LIST [G])
  require
    acyclic (list)
  local
    reversed, temp: LIST [G]
  do
    from reversed := Void
    invariant
      rev (list) catOp reversed = rev (old list)
      acyclic (list)
      acyclic (reversed)
    until list = Void
    loop
      temp := list.next
      list.next := reversed (*\label{ln:setnext}*)
      reversed := list (*\label{ln:setrev}*)
      list := temp (*\label{ln:setlist}*)
    variant list.count end
    list := reversed
  ensure
    list = rev (old list)
    acyclic (list)
  end
\end{lstlisting}
  \caption{Reversal of a linked list.}
  \label{code:reversal}
\end{figure}

Figure~\ref{code:reversal} shows the standard implementation.
Figure~\ref{fig:reversal} pictures instead a graphical representation
of \ec{reverse}'s behavior: Figure~\ref{fig:reversal-before} shows the
state of the lists in the middle of the computation, and
Figure~\ref{fig:reversal-after} shows how the state changes after one
iteration, in a way that the invariant is preserved.

The correctness proof relies on some straightforward properties of the
\ec{rev} and \ec{catOp} functions.  Initiation follows from the
property that $s \:\ec{catOp}\: \epsilon = s$. Consecution relies on the definition \eqref{eq:rev-definition} for $n > 0$, so that:
\[
\ec{rev (list.next) catOp list.first = rev (list)} .
\]
Proving the preservation of acyclicity relies on the two properties:
\begin{align*}
\ec{acyclic} (s_1\,s_2\,\ldots\,s_n) &\quad\limpl\quad \ec{acyclic} (s_2\,\ldots\,s_n) \\
|s_1| = 1 \land \ec{acyclic} (r) &\quad\limpl\quad \ec{acyclic} (s_1 \:\ec{catOp}\: r) .
\end{align*}

\subsubsection{Binary search trees}
\label{sec:bst}

Each node in a binary tree has at most two children, conventionally called \emph{left} and \emph{right}.
Binary \emph{search} trees are a special kind of binary trees whose nodes store values from some totally ordered domain \ec{T} and are arranged reflecting the relative order of their values; namely, if $t$ is a binary search tree and $n \in t$ is one of its nodes with value $v$, all nodes in $n$'s left subtree store values less than or equal to $v$, and all nodes in $n$'s right subtree store values greater than or equal to $v$.
We express this characterizing property using domain-theory notation as:
\begin{equation}
\begin{array}{rl}
s \in \ec{t[n.left]} &\quad\limpl\quad \ec{s.value} \leq \ec{n.value} \\
s \in \ec{t[n.right]} &\quad\limpl\quad \ec{s.value} \geq \ec{n.value} \,,
\end{array}
\label{eq:bst-characterization}
\end{equation}
where \ec{t[n]} denotes $t$'s subtree rooted at node $n$.
This property underpins the correctness of algorithms for operations such as searching, inserting, and removing nodes in binary search trees that run in time linear in a tree's \emph{height}; for trees whose nodes are properly balanced, the height is logarithmic in the number of nodes, and hence the operations can be performed efficiently. We now illustrate two of these algorithms with their invariants.

Consider searching for a node with value \ec{key} in a binary search tree \ec{t}.
If \ec{t.values} denotes the set of values stored in \ec{t}, the specification of this operation consists of the postcondition
\begin{align}
\ec{key} \in \ec{t.values} 
&\quad\limpl\quad  \ec{Result} \in \ec{t} \ \land\ \ec{key} = \ec{Result.value} \label{post:bst-found} \\
\ec{key} \not\in \ec{t.values}
&\quad\limpl\quad  \ec{Result = Void} \,, \label{post:bst-not-found}
\end{align}
where \ec{Void} is returned if no node has value \ec{key}.
For simplicity, we only consider non-empty trees---handling the special case of an empty tree is straightforward.

We can obtain the essential invariant by weakening both conjuncts in \eqref{post:bst-found} based on two immediate properties of trees.
First, \ec{Result $\:\in\:$ t} implies \ec{Result /= Void}, because no valid node is \ec{Void}.
Second, a node's value belongs to the set of values of the subtree rooted at the node:
\[
\ec{n.value} \;\in\; \ec{t[n].values}
\]
for $n \in t$.
Thus, the following formula is a weakening of \eqref{post:bst-found}:
\begin{equation}
\ec{key} \:\in\: \ec{t.values} \quad\limpl\quad \ec{Result /= Void} \:\land\: \ec{key} \:\in\: \ec{t[Result].values} \,,
\label{eq:bst-has-inv}
\end{equation}
which works as essential loop invariant for binary-search-tree search.

Search works by moving \ec{Result} to the left or right subtree---according to \eqref{eq:bst-characterization}---until a value \ec{key} is found or the subtree to be explored is empty.
This corresponds to the disjunctive exit condition \ec{Result = Void lor Result.value = key} and to the bounding invariant \ec{Result /= Void $\;$limplies$\;$ Result $\,\in\,$ t}: we are within the tree until we hit \ec{Void}.
Figure~\ref{code:bst-has} shows the corresponding implementation.

\begin{figure}[!htb]
\begin{lstlisting}
has_bst (t: BS_TREE [T$\,$]; key: T): NODE
  require
    t.root /= Void -- nonempty tree
  do
    from Result := t.root
    invariant
      Result /= Void $\;$limplies$\;$ Result $\in$ t
      key $\in$ t.values $\;$limplies$\;$ Result /= Void $\:$land$\:$ key $\in$ t[Result].values
    until Result = Void lor key = Result.value
    loop
      if key < Result.value then
        Result := Result.left
      else
        Result := Result.right
      end
    end
  ensure
    key $\in$ t.values $\;$limplies$\;$ Result $\in$ t land key = Result.value
    key $\not\in$ t.values $\;$limplies$\;$ Result = Void
  end
\end{lstlisting}
  \caption{Search in a binary search tree.}
  \label{code:bst-has}
\end{figure}

Initiation follows from the precondition and from the identity \ec{t[t.root] = t}.
Consecution relies on the following domain property, which in turn follows from \eqref{eq:bst-characterization}: 
\begin{align*}
n \in t
\ \land\ 
\ec{v} \in \ec{t[n].values} 
\ \land\ 
\ec{v} < \ec{n.value}
&\ \ \limpl\ \ 
\ec{n.left /= Void}
\ \land\ 
\ec{v} \in \ec{t[n.left].values}
\\
n \in t
\ \land\ 
\ec{v} \in \ec{t[n].values} 
\ \land\ 
\ec{v} > \ec{n.value}
&\ \ \limpl\ \ 
\ec{n.right /= Void}
\ \land\ 
\ec{v} \in \ec{t[n.right].values} \,.
\end{align*}

The ordering property \eqref{eq:bst-characterization} entails that the leftmost node in a (non-empty) tree $t$---that is the first node without left child reached by always going left from the root---stores the minimum of all node values.
This property, expressible using the domain-theory function \ec{leftmost} as:
\begin{equation}
\min(\ec{t.values}) = \ec{leftmost} (t).\ec{value} \,,
\label{eq:bst-leftmost-is-min}
\end{equation}
leads to an algorithm to determine the node with minimum value in a binary search tree, whose postcondition is thus:
\begin{align}
\ec{Result} &\ =\ \ec{leftmost} (t)  \label{post:bst-leftmost} \\
\ec{Result.value} &\ =\ \min (\ec{t.values}) \label{post:bst-min} \,.
\end{align}
The algorithm only has to establish \eqref{post:bst-leftmost}, which then implies \eqref{post:bst-min} combined with the property \eqref{eq:bst-leftmost-is-min}.
In fact, the algorithm is oblivious of \eqref{eq:bst-characterization} and operates solely based on structural properties of binary trees; \eqref{post:bst-min} follows as an afterthought.

Duplicating the right-hand side of \eqref{post:bst-leftmost}, writing \ec{t} in the equivalent form \ec{t[t.root]}, and applying constant relaxation to \ec{t.root} yields the essential invariant
\begin{equation}
\ec{leftmost} (\ec{t[Result]}) = \ec{leftmost} (\ec{t})
\label{eq:bst-min-essential}
\end{equation}
with the bounding invariant \ec{Result $\,\in\,$ t} that we remain inside the tree \ec{t}.
These invariants capture the procedure informally highlighted above: walk down the left children until you reach the leftmost node.
The corresponding implementation is in Figure~\ref{code:bst-minimum}.

\begin{figure}[!htb]
\begin{lstlisting}
min_bst (t: BS_TREE [T$\,$]): NODE
  require
    t.root /= Void -- nonempty tree
  do
    from Result := t.root
    invariant
      Result $\in$ t
      leftmost (t[Result]) = leftmost (t)
    until Result.left = Void
    loop
      Result := Result.left
    end
  ensure
    Result = leftmost (t)
    Result.value = min (t.values)
  end
\end{lstlisting}
  \caption{Minimum in a binary search tree.}
  \label{code:bst-minimum}
\end{figure}

Initiation follows by trivial identities.
Consecution relies on a structural property of the leftmost node in any binary tree: 
\[
\ec{n} \in \ec{t} \ \land\ \ec{n.left /= Void} 
\quad\limpl\quad
\ec{n.left} \in t \ \land\ \ec{leftmost} (\ec{t[n]}) = \ec{leftmost} (\ec{t[n.left]}) .
\]


\subsection{Fixpoint algorithms: PageRank} 
\label{sec:fixpoint-algorithms}
\label{sec:pagerank}


PageRank is a measure of the popularity of nodes in a network, used by
the Google Internet search engine. The basic idea is that the PageRank
score of a node is higher the more nodes link to it (multiple links
from the same page or self-links are ignored). More precisely, the
PageRank score is the probability that a random visit on the graph
(with uniform probability on the outgoing links) reaches it at some
point in time. The score also takes into account a \emph{dampening}
factor, which limits the number of consecutive links followed in a
random visit. If the graph is modeled by an adjacency matrix (modified
to take into account multiple and self-links, and to make sure that
there are no sink nodes without outgoing edges), the PageRank scores
are the entries of the \emph{dominant eigenvector} of the matrix.

In our presentation, the algorithm does not deal directly with the
adjacency matrix but inputs information about the graph through
arguments \ec{reaching} and \ec{outbound}. The former is an array of
sets of nodes: \ec{reaching [k]} denotes the set of nodes that
directly link to node \ec{k}. The other argument \ec{outbound [k]}
denotes instead the number of outbound links (to different nodes)
originating in node \ec{k}. The \ec{Result} is a vector of $n$ real
numbers, encoded as an array, where $n$ is the number of nodes. If
\ec{eigenvector} denotes the dominant eigenvector (also of length $n$)
of the adjacency matrix (defined implicitly), the postcondition states
that the algorithm computes the dominant eigenvector to within
precision $\epsilon$ (another input):
\begin{equation}
| \ec{eigenvector} - \ec{Result} | < \epsilon \,.
\label{eq:rank-post}
\end{equation}
That is, \ec{Result [k]} is the rank of node $k$ to within overall
precision $\epsilon$.

The algorithm computes the PageRank score iteratively: it starts
assuming a uniform probability on the $n$ nodes, and then it updates
it until convergence to a fixpoint. Before every iteration, the
algorithm saves the previous values of \ec{Result} as \ec{old_rank},
so that it can evaluate the progress made after the iteration by
comparing the sum \ec{diff} of all pairwise absolute differences, one
for each node, between the scores saved in \ec{old_rank} and the newly
computed scores available in \ec{Result}. Correspondingly, the main
loop's essential invariant is postcondition
\eqref{eq:rank-post} with \ec{diff} substituted for $\epsilon$.
\ec{diff} gets smaller with every iteration of the main loop, until it
becomes less than $\epsilon$, the main loop terminates, and the
postcondition holds. The connection between the main loop invariants and
the postcondition is thus straightforward.

\begin{figure}[!htb]
\begin{lstlisting}
page_rank (dampening, $\epsilon$: REAL; reaching: ARRAY [SET [INTEGER]]; 
           outbound: ARRAY [INTEGER]): ARRAY [REAL]
  require
    0 < dampening < 1
    $\epsilon$ > 0
    reaching.count = outbound.count = n > 0
  local
    diff: REAL
    old_rank: ARRAY [REAL]
    link_to: SET [INTEGER]
  do
    old_rank := $\{1/n\}^{n}$ -- Initialized with $n$ elements all equal to $1/n$
    from diff := 1
    invariant
      | eigenvector - Result | < diff
    until diff < $\epsilon$
    loop
      diff := 0
      across [1..n] as i loop
        Result [i] := 0
        link_to := reaching [i]
        across [1..link_to.count] as j loop
          Result [i] := Result [i] + old_rank [j] / outbound [j]
        end
        Result [i] := dampening * Result [i] + (1 - dampening)/n
        diff := diff + |Result [i] - old_rank [i]|
      end
      old_rank := Result -- Copy values of (*\textbf{Result}*) into (*\textit{old\_rank}*)
    variant 1 + diff - $\epsilon$
    end
  ensure
    |eigenvector - Result | < $\;\epsilon$
  end
\end{lstlisting}
  \caption{PageRank fixpoint algorithm.}
  \label{code:pagerank}
\end{figure}

Figure~\ref{code:pagerank} shows an implementation of this
algorithm. The two across loops nested within the main loop update the
PageRank scores in \ec{Result}. Every iteration of the
outer \ec{across} loop updates the value of \ec{Result [i]} for node
$i$ as:
\begin{equation}
\frac{(1 - \ec{dampening})}{n} +
\ec{dampening} \cdot \sum_{j} \frac{\ec{old_rank}[j]}{\ec{outbound}[j]} \,.
\label{eq:rank-result-update}
\end{equation}
The inner loop computes the sum in \eqref{eq:rank-result-update} for
$j$ ranging over the set \ec{reaching [i]} of nodes that directly
reach $i$. The invariants of the across loops express the
progress in the computation of \eqref{eq:rank-result-update}; we do
not write them down explicitly as they are not particularly
insightful from the perspective of connecting postconditions and loop
invariants.

\section{Related work: Automatic invariant inference} \label{sec:relat-work:-autom}

The amount of research work on the automated inference of invariants
is substantial and spread over more than three decades; this reflects
the cardinal role that invariants play in the formal analysis and
verification of programs.  This section outlines a few fundamental
approaches that emerged, without any pretense of being exhaustive.
A natural classification of the methods to infer invariants is between
\emph{static} and \emph{dynamic}.  Static methods
(Section~\ref{sec:static-methods}) use only the program text, whereas
dynamic methods (Section~\ref{sec:dynamic-methods}) summarize the
properties of many program executions with different inputs.

\paragraph{Program construction} 
\label{sec:program-construction}
In the introductory sections, we already mentioned classical formal
methods for program construction \cite{Dij76,Gri81,BM80,Morgan94} on
which this survey paper is based.  In particular, the connection
between a loop's postcondition and its invariant buttresses the
classic methods of program construction; this survey paper has
demonstrated it on a variety of examples.  In previous
work~\cite{FM10-YG70-post}, we developed \ginpink, a tool that
practically exploits the connection between postconditions and
invariants. Given a program annotated with postconditions, \ginpink
systematically generates mutations based on the heuristics of
Section~\ref{sec:classify-gen-technique}, and then uses the Boogie
program verifier~\cite{Lei08-Boogie2} to check which mutations are
correct invariants. The \ginpink approach borrows ideas from both
static and dynamic methods for invariant inference: it is only based
on the program text (and specification) as the former, but it
generates ``candidate'' invariants to be checked---like dynamic methods
do.

\paragraph{Reasoning about domain theories}
To bridge the gap between the levels of abstraction of domain theories
and of their underlying atomic assertions (see
Section~\ref{sec:domain-theory}), one needs to reason about first- or
even higher-logic formulas often involving interpreted theories such
as arithmetic.  The research in this area of automated theorem proving
is huge; interested readers are referred to the many reference
publications on the topic \cite{HAR,Buchberger06,BM07-book,KBook}.

\subsection{Static methods} \label{sec:static-methods}

Historically, the earliest methods for invariant inference where
\emph{static} as in the pioneering work of Karr \citeyear{Kar76}.
Abstract interpretation and the constraint-based approach are the two
most widespread frameworks for static invariant inference (see also
Bradley and Manna~\citeyear[Chap.~12]{BM07-book}).  Jhala and Majumdar~\citeyear{JhalaM09}
provide an overview of the most important static techniques and
discuss how they are applied in combination with different problems of
program verification.

\emph{Abstract interpretation} is a symbolic execution of programs
over abstract domains that over-approximates the semantics of loop
iteration.  Since the seminal work by Cousot and
Cousot~\citeyear{CC77}, the technique has been updated and extended to
deal with features of modern programming languages such as
object-orientation and heap memory-management (e.g.,
Logozzo~\citeyear{Log04} and Chang and Leino~\citeyear{CL05}). One of
the main successes of abstract interpretation has been the development
of sound but incomplete tools \cite{BCC+03-Astree} that can verify the
absence of simple and common programming errors such as division by
zero or void dereferencing.

\emph{Constraint-based} techniques rely on sophisticated decision
procedures over non-trivial mathematical domains (such as polynomials
or convex polyhedra) to represent concisely the semantics of loops
with respect to certain template properties.

Static methods are sound and often complete with respect to the class
of invariants that they can infer.  Soundness and completeness are
achieved by leveraging the decidability of the underlying mathematical
domains they represent; this implies that the extension of these
techniques to new classes of properties is often limited by
undecidability.  State-of-the-art static techniques can infer
invariants in the form of mathematical domains such as linear
inequalities \cite{CH78,CSS03}, polynomials \cite{SSM04,RCK07},
restricted properties of arrays \cite{BMS06,BHIKV09,HHKV10}, and
linear arithmetic with uninterpreted functions \cite{BHMR07}.

Following Section~\ref{sec:classify-role}, the loop invariants that
static techniques can easily infer are often a form of ``bounding''
invariant. This suggests that the specialized static techniques for
loop invariant inference discussed in this section, and the idea of
deriving the loop invariant from the postcondition, demonstrated in
the rest of the paper, can be fruitfully combined: the former can
easily provide bounding loop invariants, whereas the latter can
suggest the ``essential'' components that directly connect to the
postcondition.

To our knowledge, there are only a few approaches to static invariant
inference that take advantage of existing annotations
\cite{PV04-spin,Jan07,dCGG09,LQGVW09,KV09}. 
Janota~\citeyear{Jan07} relies on user-provided assertions nested within
loop bodies and tries to check whether they hold as invariants of the
loop. The approach has been evaluated only on a limited number of
straightforward examples.
De Caso et al.~\citeyear{dCGG09} briefly discuss deriving the invariant of
a ``for'' loop from its postcondition, within a framework for
reasoning about programs written in a specialized programming
language.  Lahiri et al.~\citeyear{LQGVW09} also leverage specifications
to derive intermediate assertions, but focusing on lower-level and
type-like properties of pointers.  On the other hand, P{\u a}s{\u
  a}reanu and Visser~\citeyear{PV04-spin} derive candidate invariants
from postconditions within a framework for symbolic execution and
model-checking.

Finally, Kov{\'a}cs and Voronkov~\citeyear{KV09} derive complex loop
invariants by first encoding the loop semantics as recurring relations
and then instructing a rewrite-based theorem prover to try to remove
the dependency on the iterator variables in the relations. This
approach exploits heuristics that, while do not guarantee
completeness, are practically effective to derive automatically loop
invariants with complex quantification---a scenario that is beyond
the capabilities of most other methods.

\subsection{Dynamic methods} \label{sec:dynamic-methods} 

Only in the last decade have dynamic techniques been applied to
invariant inference.  The Daikon approach of Ernst et
al.~\citeyear{ECGN01} showed that dynamic inference is practical and
sprung much derivative work (e.g., Perkings and Ernst~\citeyear{PE04},
Csallner et al.~\citeyear{CTS08}, Polikarpova et al.~\citeyear{PCM09},
Ghezzi et al.~\citeyear{GhezziMM09}, Wei et
al.~\citeyear{WFKM11-ICSE11}, Nguyen et al.~\citeyear{NguyenKWF12},
and many others).  In a nutshell, the dynamic approach consists in
testing a large number of candidate properties against several program
runs; the properties that are not violated in any of the runs are
retained as ``likely'' invariants.  This implies that the inference is
not sound but only an ``educated guess'': dynamic invariant inference
is to static inference what testing is to program proofs.
Nonetheless, just like testing is quite effective and useful in
practice, dynamic invariant inference can work well if properly
implemented. With the latest improvements \cite{WRFPHSNM-ASE11},
dynamic invariant inference can attain soundness of over 99\% for the
``guessed'' invariants.

Among the numerous attempts to improve the effectiveness and
flexibility of dynamic inference, Gupta and
Heidepriem~\citeyear{GuptaH03} suggest to improve the quality of
inferred contracts by using different test suites (based on code
coverage and invariant coverage), and by retaining only the contracts
that are inferred with both techniques.  Fraser and
Zeller~\citeyear{fraserICST11} simplify and improve test cases based
on mining recurring usage patterns in code bases; the simplified tests
are easier to understand and focus on common usage.  Other approaches
to improve the quality of inferred contracts combine static and
dynamic techniques \cite{CTS08,TillmannCS06,WRFPHSNM-ASE11}.

To date, dynamic invariant inference has been mostly used to infer pre- 
and postconditions or intermediate assertions, whereas it has been
only rarely applied~\cite{NguyenKWF12} to \emph{loop invariant}
inference. This is probably because dynamic techniques require a
sufficiently varied collection of test cases, and this is more
difficult to achieve for loops---especially if they manipulate
complex data structures.

\section{Lessons from the mechanical proofs} \label{sec:boogie-assess}

Our verified Boogie implementations of the algorithms mirror the
presentation in Section~\ref{sec:catalog}, as they introduce the
predicates, functions, and properties of the domain theory that are
directly used in the specification and in the correctness proofs. The
technology of automatic theorem proving is, however, still in active
development, and faces in any case insurmountable theoretical limits
of undecidability. As a consequence, in a few cases we had to
introduce explicitly some intermediate domain properties that were, in
principle, logical consequences of other definitions, but that the
prover could not derive without suggestion. Similarly, in other cases
we had to guide the proof by writing down explicitly some of the
intermediate steps in a form acceptable to the verifier.

A lesson of this effort is that the syntactic form in
which definitions and properties of the domain theory are expressed
may influence the amount of additional annotations required for proof
automation. The case of the sorting algorithms in
Section~\ref{sec:sorting} is instructive. All
rely on the definition of predicate \ec{sorted} as:
\begin{equation}
\forall i \in \ec{[a.lower..a.upper - 1]: a [i] <= a [i + 1]} ,
\label{eq:sorted-conclusion}
\end{equation}
which compares \emph{adjacent} elements in positions $i$ and
$i+1$. Since bubble sort rearranges elements in an array also by
swapping adjacent elements, proving it in Boogie was straightforward,
since the prover could figure out how to apply definition
\eqref{eq:sorted-conclusion} to discharge the various verification
conditions---which also feature comparisons of adjacent elements.
Selection sort and insertion sort required substantially more guidance
in the form of additional domain theory properties and detailing of
intermediate verification steps, since their logic does not operate
directly on adjacent elements, and hence their verification conditions
are syntactically dissimilar to \eqref{eq:sorted-conclusion}. Changing
the definition of \ec{sorted} into something that relates non-adjacent
elements---such as $\forall i, j: \ec{a.lower} \leq i \leq j \leq
\ec{a.upper} \limpl a[i] \leq a[j]$---is not sufficient to bridge the
semantic gap between sortedness and other predicates: the logic of
selection sort and insertion sort remains more complex to reason about
than that of bubble sort.  On the contrary, comb sort was as easy as
bubble sort, because it relies on a generalization of \ec{sorted} that
directly matches its logic (see Section~\ref{sec:comb-sort}).

A similar experience occurred for the two dynamic programming
algorithms of Section~\ref{sec:dynamic}. While the implementation of
Levenshtein distance closely mirrors the recursive definition of
\ec{distance} (Section~\ref{sec:levenshtein}) in computing the
minimum, the relation between specification and implementation is less
straightforward for the knapsack problem (Section~\ref{sec:knapsack}),
which correspondingly required a more complicated axiomatization for the 
proof in Boogie.

\section{Conclusions and assessment} \label{sec:conclusions}

The concept of loop invariant is, as this review has attempted to
show, one of the foundational ideas of software construction. We have
seen many examples of the richness and diversity of practical loop
invariants, and how they illuminate important algorithms from many
different areas of computer science. We hope that these examples
establish the claim, made at the start of the article, that the
invariant is the key to every loop: to devise a new loop so that it is
correct requires summoning the proper invariant; to understand an
existing loop requires understanding its invariant.

Invariants belong to a number of categories, for which this discussion
has established a classification which we hope readers will find
widely applicable, and useful in understanding the loop invariants of
algorithms in their own fields. The classification is surprisingly
simple; perhaps other researchers will find new criteria that have
eluded us.

Descriptions of algorithms in articles and textbooks has increasingly,
in recent years, included loop invariants, at least informally; we
hope that the present discussion will help reinforce this trend and
increase the awareness---not only in the research community but also
among software practitioners---of the centrality of loop invariants in
programming.

Informal invariants are good, but being able to express them formally
is even better. Reaching this goal and, more generally, continuing to
make invariants ever more mainstream in software development requires
convenient, clear and expressive notations for expressing loop
invariants and other program assertions. One of the contributions of
this article will be, we hope, to establish the practicality of
domain-theory-based invariants, which express properties at a high
level of abstraction, and their superiority over approaches that
always revert to the lowest level (suggesting a possible slogan:
``Quantifiers considered harmful'').

Methodological injunctions are necessary, but the history of software
practice shows that they only succeed when supported by effective
tools. Many programmers still find it hard to come up with invariants,
and this survey shows that they have some justifications: even though
the basic idea is often clear, coming up with a sound and complete
invariant is an arduous task. Progress in invariant inference, both
theoretical and on the engineering side, remains essential. There is
already, as noted, considerable work on this topic, often aimed at
inferring more general invariant properties than the inductive loop
invariants of the present discussion; but much remains to be done to
bring the tools to a level where they can be integrated in a standard
development environment and routinely suggest invariants, conveniently
and correctly, whenever a programmer writes a loop. The work mentioned
in Section~\ref{sec:relat-work:-autom} is a step in this direction.

Another major lessons for us from preparing this survey (and
reflecting on how different it is from what could have been written on
the same topic 30, 20, 10 or even 5 years ago) came from our success
in running essentially all the examples through formal, mechanized
proofs. Verification tools such as Boogie, while still a work in
progress, have now reached a level of quality and practicality that
enables them to provide resounding guarantees for work that, in the
past, would have remained subject to human errors.

We hope that others will continue this work, both on the conceptual
side---by providing further insights into the concept of loop
invariant---and on the practical side---by extending the concept to
its counterpart for data (i.e., the class invariant) and by broadening
our exploration and classification effort to many other important
algorithms of computer science.

\paragraph{Acknowledgments}
This work is clearly based on the insights of Robert Floyd, C.A.R.\
Hoare and Edsger Dijkstra in introducing and developing the notion of
loop invariant. The mechanized proofs were made possible by Rustan
Leino's Boogie tool, a significant advance in turning axiomatic
semantics into a practically usable technique. We are grateful to our
former and present ETH colleagues Bernd Schoeller, Nadia Polikarpova,
and Julian Tschannen for pioneering the use of Boogie in our
environment.

\ifcsur
\else
\bibliographystyle{plain}
\fi





\end{document}